\documentclass[%
 aip,
 jmp,%
 amsmath,amssymb,
preprint,%
]{revtex4-1}
\usepackage{graphicx}
\usepackage{color}
\usepackage{dcolumn}
\usepackage{bm}
\usepackage{hyperref}

\input xy
\xyoption{all}

\newcommand{\be}{\begin{equation}}
\newcommand{\ee}{\end{equation}}
\newcommand{\en}{\end{equation}}

\newcommand{\ba}{\begin{array}}
\newcommand{\ea}{\end{array}}

\newcommand{\bea}{\begin{eqnarray}}
\newcommand{\ena}{\end{eqnarray}}

\newcommand{\beano}{\begin{eqnarray*}}
\newcommand{\enano}{\end{eqnarray*}}

\newcommand{\bei}{\begin{itemize}}
\newcommand{\eni}{\end{itemize}}

\newcommand{\bee}{\begin{enumerate}}
\newcommand{\ene}{\end{enumerate}}

\newtheorem{theorem}{Theorem}[section]
\newtheorem{coroll}[theorem]{Corollaire}
\newtheorem{lemma}[theorem]{Lemma}
\newtheorem{prop}[theorem]{Proposition}
\newtheorem{fit}[theorem]{Definition}

\newcommand{\betheo}{\begin{theorem}}
\newcommand{\entheo}{\end{theorem}}

\newcommand{\becor}{\begin{coroll}}
\newcommand{\encor}{\end{coroll}}

\newcommand{\belem}{\begin{lemma}}
\newcommand{\enlem}{\end{lemma}}

\newcommand{\befit}{\begin{fit}}
\newcommand{\enfit}{\end{fit}}

\newcommand{\beprop}{\begin{prop}}
\newcommand{\enprop}{\end{prop}}

\def \k0 {\frac{1 }{4 \pi \epsilon_0}}

\def \bc {\begin{center}}
\def \ec {\end{center}}

\def\half{\mbox{\small{$\frac{1}{2}$}}}

\newcommand{\sech}{{\textrm{sech}}}

\def \pt {.}

\newcommand{\virg}{{\textrm{\ ,}}}

\def \dis {\displaystyle}

\def \id {\hbox{\bf \large 1} \!\! \hbox{\bf \large \sf I}}
\def \IN {\rm I \! N}

\def \IR {\rm I \! R}
\def \IC {{\sf I} \!\!\! {\rm C}}

\def  \ni {\noindent}

\def  \lv {\left|}
\def  \rv {\right|}

\def  \li {\left<}
\def  \rs {\right>}

\begin{document}


\title[]{Photon-added coherent states for shape invariant systems}

\author{Komi Sodoga }
\email{ksodoga@tg.refer.org}

\author{Mahouton Norbert Hounkonnou }
\email{ norbert.hounkonnou@cipma.uac.bj, with copy to hounkonnou@yahoo.fr}

\author{Isiaka Aremua }
\email{ claudisak@yahoo.fr}

\affiliation{
$^{a,c}$ Universit\'e de Lom\'e Facult\'e des Sciences, D\'epartement de Physique,\\
 Laboratoire de Physique des Mat\'eriaux et des Composants \`a Semi-Conducteurs,\\
 02 BP 1515 Lom\'e, Togo\\\quad \\
$^{a,b,c}$University of Abomey-Calavi,\\
International Chair in Mathematical Physics and Applications\\
(ICMPA--UNESCO Chair), 072 B.P. 50  Cotonou, Republic of Benin}

\date{\today}

\begin{abstract}
This paper addresses a full characterization of photon-added coherent states for shape-invariant potentials. Main properties are investigated and discussed. A statistical computation of relevant physical quantities is performed, emphasizing the importance of using generalized hypergeometric functions $_pF_q$ and Meijer's $G$-functions for such a study.

\end{abstract}
\maketitle

\section{Introduction}
 \ni Since the introduction of  the concept of canonical coherent states (CS), associated with the one dimensional
harmonic oscillator   by Schr\"{o}dinger in 1926 \cite{Schrodinger}, followed decades in which this concept
has reached great  investigations \cite{Klauder_App}-\cite{Gazeau}. CS  are
a useful mathematical framework for dealing with the connection between classical and
quantum mechanics \cite{Gazeau, Berezin, Aremua1}.   These states can globally be
constructed in three equivalent ways: (i) by defining them as eigenstates of the lowering operator 
(called CS of the Barut-Girardello  \cite{Barut} type), 
(ii) by applying a unitary displacement operator on a ground state (Klauder-Perelomov CS \cite{Perelomov} 
or CS of the Gazeau-Klauder type),  
and (iii) by considering them as quantum states with a minimum uncertainty
relationship \cite{Aragone, Nieto}. 

 The CS for shape-invariant potentials (SIP) performed in this work \cite{Fukui, Aleixo}, belong to the Barut-Girardello type. 
They are built using algebraic approach based on the supersymmetric quantum mechanics (SUSY-QM) \cite{Witten, Cooper}. 
SUSY QM deals with the study of partner Hamiltonians which are isospectral, that is, they have {\it almost} the same energy eigenvalues. 
A number of such partner Hamiltonians satisfy an integrability condition called shape invariance \cite{Gend}-\cite{Dab}. 
However, not  all exactly solvable
systems are shape-invariant.

 Recently, 
a considerable attention is devoted to
photon-added CS (PACS)\cite{Dodonov}-\cite{Daoud}, first introduced
by Agarwal and Tara \cite{Aga}. The PACS represent  interesting states generalizing both the Fock states and CS.  
Indeed, they are obtained by repeatedly operating the photon
  creation operator on an ordinary CS. In some previous works, the PACS were assimilated to 
  nonlinear CS.  Their various generalizations  were also performed \cite{Lietal, Zhang}. They evidence some 
  nonclassical effects, for e.g, amplitude squeezing, sub-Poissonian behaviour, nonclassical 
  quasi-probability distribution. In one of our previous papers  \cite{hounk-ngompe}, a family of photon 
  added as well as photon depleted CS related to the inverse of ladder operators acting on hypergeometric CS  was
  introduced. Their squeezing and antibunching 
  properties were investigated in both standard (nondeformed) and deformed quantum optics. 
   Recently \cite{mojaverietal1}, new generalized  PACS 
were formulated by excitations on a  family of generalized CS.  
 Their non-classical features and  quantum statistical
properties   were compared with those obtained by  Agarwal. Besides, in another paper \cite{mojaverietal2}, photon-subtracted generalized CS,  
which are reminiscent of the PACS,  were introduced;  their nonclassical features were also discussed.
PACS find many applications in physics. In   \cite{Dakna}, 
 a generating  Schr\"{o}dinger-cat-like states of a single-mode optical field scheme was provided,  leading to  properties 
 similar  to those of superpositions of two CS with relevant statistical quantities  which were analytically and numerically treated.  
 In \cite{Ban},  the photon number average value dependence, and the factor  of the conditional output state on the measurement outcome 
 as well as the 
statistical distance between the input and
conditional output states, were investigated in the study of  a lossless beam splitter.
All these applications motivate the necessity  to look at this
kind of states associated to the generalized CS for shape-invariant potentials,
 investigated by Aleixo and Balantekin\cite{Aleixo},
in a  unified description of  different CS for exactly solvable quantum systems.  

All known previous  works on PACS were based on SUSYQM factorization or similar methods generating exactly solvable potentials. Unfortunately, to our best knowledge of the literature, the SIP approach for PACS is still lacking. Our present study also aims at filling this gap.
Indeed,  we are interested  in producing the  PACS from the shape-invariant potential CS.  
These new states are denoted by {\it photon-added shape invariant potential CS} (PA-SIPCS). Their mathematical and physical properties are  defined and discussed in details. Relevant examples are explicitly treated as illustration.\\
 The paper is organized as follows. 
First, in Section 2,  a brief review of SUSY-QM factorization, the algebraic formulation of 
shape-invariance condition and the construction of the generalized shape-invariant potentials 
coherent states  (SIPCS) are provided. In  
Section 3,   PA-SIPCS are built by
successive applications of the raising operator on the SIP-CS.  The   inner product  of 
two different PA-SIPCS   is nonzero, 
evidencing  that the obtained states are not mutually orthogonal. Besides,  the   
   normalization factor is determined and the resolution of unity  studied. Next, their reproducing kernel insights, 
   due to their overcompleteness on the quantum Hilbert space property, are analyzed.  Then, their   
   statistical properties are  determined and discussed. 
   Different concrete examples, based on  Infeld and Hull \cite{Infeld} classification,  
   are   furnished in Section 4, on different types 
  of shape-invariant systems.  Finally, in Section 5,  we end with some  concluding remarks.

\section{Quick overview on SUSY-QM factorization, shape invariance and related generalized CS}
\ni For  convenience, let us consider here  a one dimensional Hamiltonian 
\bea \label{eq1}
H = - {d^2 \over dx^2}  + V(x) , \quad  x \in I \subset \IR,
\ena
where for notation simplification we set $\hbar = 2m = 1$.
The  SUSY-QM factorization of the Hamiltonian (\ref{eq1}) consists in  writing 
\bea \label{eq2}
H - E_0:= A^\dag A,
\ena
where $E_0$ is the ground state energy of $H$ corresponding to the ground state $ \Psi_0.$ The formal  mutually adjoint operators $A$ and $A^\dag$ are  defined by 
\bea \label{eq2bis}
 A :=\dis {d\over dx} + W(x)\quad \textrm{and}\quad   A^\dag  := -\dis{d\over dx} + W(x)
 \quad \textrm{with}\quad 
 W(x)=  -\dis {d\over dx}[\ln \Psi_0(x)].
\ena
The function $W$, called superpotential, related to the ground state eigenfunction,   is solution of  the Riccati equation 
$V^2(x) - E_0 = W^2(x) - W'(x)$. 
The partner Hamiltonians $H_{1,2}$  are
\bea \label{eq3}
H_1 =  -{d^2 \over dx^2} + V_1(x), \quad H_2 = -{d^2 \over dx^2} + V_2(x).
\ena
 $H_1$  is expressed through the initial Hamiltonian as:  $H_1 = H - E_0$. 
The partner potentials are defined by: 
\bea \label{eq4}
V_1(x) := W^2 (x) - W'(x),  \quad V_2(x) := W^2 (x) + W'(x) \pt
\ena
The partner Hamiltonians $H_1, H_2$ and mutually adjoint operators $A$ and $A^\dag$ are linked as: 
\bea \label{eq5}
H_1 A^\dag = A^\dag H_2,  \quad H_2 A = A H_1 \pt 
\ena
From equation (\ref{eq5}) one can show that $H_1$ and $H_2$ are isospectral, i.e,
\bea \label{eq6}
\textrm{if} \quad 
\left\{ \ba{rcl} 
H_1 \Psi_n^{(1)} & = & E_n^{(1)} \Psi_n^{(1)} \\
H_2 \Psi_n^{(2)} & = & E_n^{(2)} \Psi_n^{(2)}
\ea \right.  \quad \textrm{then} \quad 
\left\{ \ba{rcl} 
 \Psi_n^{(2)} \propto A  \Psi_{n+1}^{(1)} \\
 \Psi_{n+1}^{(1)} \propto A^\dag \Psi_n^{(2)}
\ea \right. \pt
\ena
 More precisely,  the spectra  of $H_1$ and $H_2$ are related as:
\bea  \label{eq7}
 E_n^{(2)} = E_{n+1}^{(1)}, \  \Psi_n^{(2)} = [E_{n+1}^{1}]^{-1/2} A \Psi_{n+1}^{(1)},\ 
\Psi_{n+1}^{(1)} =  [E_{n}^{2}]^{-1/2} A^\dag \Psi_{n}^{(2)},\ 
E_0^{(1)}  = 0. \  
\ena
These relations (\ref{eq7}) express that if the spectra of one of the partners, say $H_1$, are  known, one can immediately deduce the spectra of $H_2$. However, Eqs. (\ref{eq7}) only give the relations between the spectra of the two partner Hamiltonians, but do not allow to determine their spectra. 

 A criterion  of an exact solvability is known as shape invariance condition \cite{Gend}; that is the pair of SUSY partner potentials $V_{1,2}$ are similar in shape and differ only in the parameters that appear in them. 
\befit
 The SUSY partner potentials $V_{1,2}$ are said shape-invariant  if:
\bea \label{eq8}
V_2(x ; a_1) = V_1(x ; a_2) + {\cal R}(a_1)
\ena
where $a_1$ is a set of parameters: $a_2 := f(a_1)$  a functional of $a_1$, and ${\cal R}(a_1)$ is a remainder, independent  from the dynamical variables $x$ and $p_x$. 
\enfit
Constructing a hierarchy of Hamiltonians by repeated re-factorization of $H_1$, the shape invariance condition (\ref{eq8})  allows to explicitly deduce   the eigenvalues and eigenfunctions \cite{Dutt4}-\cite{Dab} as follows:
\bea \label{eq9}
  & & E_n^{(1)} := \sum_{k = 1}^n {\cal R}(a_k),\quad 
\Psi_n^{(1)}(x ; a_1)\propto A^\dag(x ; a_1) \ldots A^\dag(x ; a_n) .A^\dag(x ; a_{n+1})\, \psi_0^{(1)}(x; a_{n+1}) \pt
\ena
The method of shape-invariant potentials can be  viewed as a generalization of the operator method for harmonic oscillator. 
Then, a question  immediately arises: is there any algebraic structure underlying the shape-invariant potentials (SIP), 
similar to the Harmonic oscillator (Weyl Heisenberg) algebra?
 The answer to this question is provided by Fukui \cite{Fukui}, and Balantekin \cite{Aleixo}.
 SIP condition (\ref{eq8}) can be rewritten in operator form as:
$A(a_1) A^\dag (a_1) = A^\dag(a_2) A(a_2) + {\cal R}(a_1)$.
Introducing a similarity transformation  $T$ which replaces, in a given operator ${\cal O}(x; a_1)\equiv {\cal O}(a_1)$, $a_1$ by $a_2$:
$ T(a_1) {\cal O}(a_1) T^\dag(a_1)= {\cal O}(a_2)$, and the operators 
\bea \label{eq10}
B_+ := A^\dag(a_1) T(a_1), \quad B_- :=  T^\dag(a_1)  A(a_1) 
\ena
 the Hamiltonian can be factorized in terms of the new operators as follows:
\bea \label{eq11}
H - E_0 = H_1 = A^\dag(a_1) A(a_1) = B_+ B_-
\ena
where
\bea \label{eq12}
[B_-, B_+] & := & {\cal R}(a_0),\quad
B_-\lv \Psi_0\rs  := 0 \pt
\ena
Let us introduce the Hilbert space $\mathfrak{H}$ spanned by the states $\lv \Psi_n\rs$ given by 
\bea
\mathfrak{H} :=  span \left\{\lv \Psi_n\rs, n   = 0, 1, 2, \cdots, + \infty\right\}
\ena
in which the following relation 
\bea
\sum_{n= 0}^{\infty}\lv \Psi_n\rs  \li \Psi_n\rv  = \id_{{\mathfrak H}}
\ena
holds, where $\id_{{\mathfrak H}}$ is the identity operator  on $\mathfrak{H}.$\\
The states $B_+^n\lv\Psi_0\rs$ are eigenfunctions of $ H$ with eigenvalues $E_n$, that is:
\bea \label{eq13}
H(B_+^n \lv \Psi_0\rs) & = & \underbrace{\left[\sum_{k = 1}^n {\cal R}(a_k)\right]}_{E_n} B_+^n \lv \Psi_0\rs.
\ena
$B_{\pm}$  act as raising and lowering operators: 
\bea \label{eq14}
B_+\lv \Psi_n\rs  & := & \sqrt{E_{n+1}} \lv\Psi_{n+1}\rs, \quad 
B_-\lv \Psi_n\rs := \sqrt{ {\cal R}(a_0) + E_{n-1}} \lv \Psi_{n-1}\rs \pt
\ena

 To define the  shape-invariant potential coherent states (SIPCS), Balantekin {\it etal} \cite{Aleixo} introduce 
 the right inverse $B_-^{-1}$ of $B_-$ as: $B_- B_-^{-1}  = \id$
and the left inverse $H^{-1}$ of $H$ such that: $H^{-1} B_+ := B_-^{-1}$. 
The SIPCS defined by 
\bea \label{eq15}
\lv z\rs = \sum_{n = 0}^n (z B_-^{-1})^n \lv\Psi_0\rs
\ena
are   eigenstates of the lowering operator $B_-$: 
\bea \label{eq16}
B_-\lv z\rs & = & z \lv z\rs\pt
\ena
A generalization of the SIPCS (\ref{eq15}) is performed in \cite{Aleixo}: 
\bea \label{eq17}
\lv z ; a_j\rs = \sum_{n=0}^\infty \left\{ z {\cal Z}_j B_-^{-1}\right\}^n \lv \Psi_0\rs, \quad z,{\cal Z}_j \in \IC
\ena
where  ${\cal Z}_j \equiv {\cal Z}(a_j) = {\cal Z}(a_1, a_2, \ldots) $.\ 
Observing that  $B_-^{-1}{\cal Z}_j = {\cal Z}_{j+1}\ B_-^{-1}$,
and from 
\bea \label{eq18}
{\cal Z}_{j-1}  := T^\dag(a_1) {\cal Z}_j T(a_1),
\ena
one can readily show that 
\bea \label{eq19}
(z {\cal Z}_j B_-^{-1})^n = z^n \prod_{k = 0}^{n - 1} {\cal Z}_{j+k} B_{-}^{-n}\pt
\ena
Using the relation  (\ref{eq18}), one can straightforwardly  show that the states  (\ref{eq17}) are  eigenstates of $B_-$: 
\bea \label{eq20}
B_- \lv z; a_j\rs = z {\cal Z}_{j -1} \lv z;a_j\rs \pt
\ena
Observing that 
\bea \label{eq21}
B_-^{-n}\lv \Psi_0\rs = C_n\lv\Psi_n\rs , \quad C_n = \left[\prod_{k = 1}^n\left(\sum_{s = k}^n {\cal R}(a_s) \right) \right]^{-1/2}
\ena
and using  Eq. (\ref{eq19}), 
the
 normalized form of the CS (\ref{eq17}) can be obtained as: 
\bea \label{eq22}
\lv z; a_r\rs = {\cal N}(|z|^2; a_r)\sum_{n = 0}^\infty {z^n \over h_n(a_r)}\lv \Psi_n\rs 
\ena
where we used the shorthand notation $a_r := [{\cal R}(a_1), {\cal R}(a_2), \ldots,{\cal R}(a_n)\,;\,a_j, a_{j+1},\ldots, a_{j+n -1} ]$. 
The expansion coefficient $h_n(a_r)$ and the normalization constant $ {\cal N}(|z|^2; a_r)$ are:
{\small
\bea \label{eq23}
 h_n(a_r) = \dis\frac{\sqrt{\dis \prod_{k = 1}^n \left(\sum_{s = k}^n {\cal R}(a_s) \right)}}
{\dis \prod_{k = 0}^{n - 1} Z_{j+k}} \quad 
\textrm{for} \  n \ge 1,\   h_0(a_r) = 1, 
\quad{\cal N}(x; a_r) = \left[\sum_{n = 0}^\infty \, {x^n \over |h_n(a_r)|^2 } \right]^{-1/2}\pt
\ena
}
It is shown that these states (\ref{eq22}) fulfill the standard properties of label continuity, overcompleteness, temporal stability, and action identity \cite{Aleixo}.

\section{Construction of photon-added coherent states for shape invariant systems }
\subsection{Definition of the PA-SIPCS}
\ni Let  ${\mathfrak H}_m $  be the  Hilbert subspace  of $\mathfrak H$ defined  as follows
\bea
 {\mathfrak H}_m := span\left\{\lv\Psi_{n+m}\rs\right\}_{n,m \ge 0}.
\ena
By repeated applications of the raising operator $B_+$ on  the generalized SIPCS 
 (\ref{eq17}), we can obtain 
photon-added shape-invariant potential CS (PA-SIPCS)  denoted by  $\lv z ; a_r \rs_m$  as follows:
\bea \label{eq24}
\lv z ; a_r \rs_m & :=  & (B_+^m)\lv z ; a_r \rs
\ena
 where $m$ is a positive integer standing for  the number of added quanta or  photons. \\
It is worth mentioning that  the first $m$ 
eigenstates $\lv\Psi_n\rs $, $n =0,1, \dots, m-1$ are absent from the wavefunction $\lv z ; a_r \rs_m \in {\mathfrak H}_m.$  
Therefore, from the 
orthonormality relation satisfied by the states $\lv\Psi_n\rs,$   the  overcompleteness relation fulfilled by 
the  identity operator on  ${\mathfrak H}_m,$ denoted  by 
$\id_{{\mathfrak H}_m}$, is to  be written as \cite{Penson, Popov}
\bea
 \id_{{\mathfrak H}_m} = \sum_{n = m}^\infty \lv\Psi_{n} \rs \li \Psi_{n}\rv = \sum_{n = 0}^\infty \lv\Psi_{n+m} \rs \li \Psi_{n + m}\rv.
\ena 
Here, $\id_{{\mathfrak H}_{m}}$ is only required to be a bounded positive operator with a densely defined inverse 
\cite{Ali95}.
\ni From (\ref{eq23}),  and using the relation $B_+ {\cal R}(a_{n-1}) = {\cal R}(a_n) B_+$,  we obtain:
\bea \label{eq25}
B_+ {1\over h_n(a_r)} = \dis \frac{\dis \prod_{k = 1}^{n} {\cal Z}_{j+k}}
{\dis \left[\prod_{k = 2}^{n+1} \left(\sum_{s = k}^n {\cal R}(a_s) \right) \right]^{1/2}  } B_+ \pt
\ena
From Eq. (\ref{eq14}), $B_+\lv \Psi_n\rs = \sqrt{{\cal R}(a_1) + \ldots + {\cal R}(a_{n+1})} \lv \Psi_{n+1}\rs$,
we obtain:
\bea \label{eq26}
B_+{z^n \over h_n(a_r)} \lv \Psi_n\rs =
\dis \frac{\dis  \left(\prod_{k = 1}^{n} Z_{j+k}\right).\left(\sum_{s=1}^{n+1}{\cal R}(a_s) \right)^{1/2}     } 
{ \dis \left[\prod_{k = 2}^{n+1} \left(\sum_{s = k}^n {\cal R}(a_s) \right) \right]^{1/2} } z^n\lv \Psi_{n+1}\rs\pt
\ena
Repeated applications of $B_+$ give:
\bea \label{eq27}
B_+^m {z^n \over h_n(a_r)} \lv \Psi_{n}\rs = { z^n \over K_n^m(a_r)} \lv \Psi_{n+m}\rs \virg
\ena
where the expansion coefficient takes the form:
\bea \label{eq28}
K_n^m(a_r)  & = & \dis \frac{ \dis \left[\prod_{k = m+1}^{n+m} \left(\sum_{s = k}^{n+m} {\cal R}(a_s) \right) \right]^{1/2} }
{\dis  \left[\prod_{k = m}^{n+m-1} Z_{j+k}\right].
\left[ \prod_{k = 1}^m \left(\sum_{s=k}^{n+m}{\cal R}(a_s)\right) \right]^{1/2}} 
 \pt
\ena
Then, the  PA-SIPCS can be written as:
\bea  \label{eq29}
\lv z; a_r\rs_m = B_+^m \lv z; a_r\rs = \sum_{n = 0}^\infty {z^n \over K_n^m(a_r) } \lv \Psi_{n+m}\rs \pt 
\ena
\subsection{Normalization and non-orthogonality}
 \ni We can obtain the normalized form of the PA-SIPCS 
\bea \label{eq30}
\lv z;a_r\rs_m = {\cal N}_m(|z|^2; a_r) \sum_{n = 0}^\infty {z^n \over K_n^m(a_r) } \lv \Psi_{n+m}\rs
\ena
by requiring that \ ${}_m\li z;a_r\right.\lv z; a_r\rs_m = 1$. The normalization constant ${\cal N}_m(|z|^2; a_r)$ is given by:
\bea \label{eq31}
{\cal N}_m(|z|^2; a_r) = \left( \sum_{n = 0}^\infty {|z|^{2n} \over |K_n^m(a_r)|^2 }\right)^{-1/2} \pt
\ena

 The inner product of two different PA-SIPCS $\lv z; a_r\rs_m$ and $\lv z'; a_r\rs_{m'}$ 
\bea \label{eq32}
 {}_{m'}\li z';a_r\right.\lv z; a_r\rs_m & = &  {\cal N}_{m'}(|z'|^2; a_r) {\cal N}_m(|z|^2; a_r) 
\sum_{n,n' = 0}^\infty { {{z'}^\star}^{n'} z^n  \over {K_{n'}^{m'}}^\star(a_r) K_n^m(a_r)} %
\li \Psi_{n'+m'} \right. \lv  \Psi_{n+m} \rs
\ena 
does not vanish. Indeed, due to the orthonormality of the eigenstates $|\Psi_n\rangle$, the inner product (\ref{eq32}) can be rewritten as
\bea \label{eq33}
 {}_{_{m'}}\li z';a_r\right.\lv z; a_r\rs_m & = &  {\cal N}_{m'}(|z'|^2; a_r) {\cal N}_m(|z|^2; a_r) {z'^\star}^{(m - m')}
\sum_{n = 0}^\infty {({z'}^\star z)^n \over  {K_{n +m-m'}^{m'}}^\star(a_r)\ K_n^m(a_r)},
\ena
 showing that the PA-SIPCS are not mutually orthogonal. 
 
\subsection{Overcompleteness}
\ni  We assume the existence of a non-negative weight function $\omega_m $ such that the overcompleteness  or the resolution of identity 
\bea \label{eq34}
 \int_{{\IC}}\  d^2z\, \lv z; a_r\rs_m\, \omega_m(|z|^2;a_r)\, {}_{m}\li z;a_r\rv  = \id_{{\mathfrak H}_m} 
\ena
holds.

Inserting the definition (\ref{eq30}) of the PA-SIPCS $\lv z; a_r\rs_m$ into Eq. (\ref{eq34}) yields: 
\bea \label{eq35}
 \int_{{\IC}}\  d^2z\  {\cal N}^2_m(|z|^2; a_r) \sum_{n,n' =0}^\infty 
{ {z^\star}^{n'}z^n  \over   {K_{n'}^m}^\star(a_r) K_n^m(a_r)}\ 
  \lv \Psi_{n' + m} \rs \li \Psi_{n + m} \rv \omega_m(|z|^2;a_r) = \id_{{\mathfrak H}_m}\pt
\ena
The diagonal matrix elements of the above relation, using the orthonormality of the eigenfunctions 
$\lv \Psi_n\rs$, gives:
\bea \label{eq36}
 \int_{\IC} \ d^2z\  {\cal N}^2_m(|z|^2; a_r) |z|^{2n} \omega_m(|z|^2; a_r) =  |K_n^m(a_r)|^2 \pt
\ena
We can see, after  computation of the angular integration,
  that the weight function $\omega_m$ must fulfill  the condition:
\bea \label{eq37}
 \int_0^\infty dx\  x^n\  {\cal W}_m(x ; a_r) =  |K_n^m(a_r)|^2, \quad 
\quad \textrm{where} \quad 
 {\cal W}_m(x ; a_r) = \pi {\cal N}_m^2(x; a_r)\  \omega_m(x ; a_r)\pt
\ena
Here we use  the polar  representation $z = r e^{i\phi}$; $x$ stands for $|z|^2 = r^2$. 
Therefore, the weight function $\omega_m$ is related to the undetermined  moment distribution ${\cal W}_m(x ; a_r)$, which is the  solution of the Stieltjes moment problem with the moments given by $ |K_n^m(a_r)|^2$.\\
Let us point out here that there are several methods to determine the measure $\omega_m(x; a_r)$. 
Depending on the form of the coefficient  $K_n^m(a_r)$, we can refer to well known  results following  standard handbooks of tabulated integrals
\cite{Marichev}-\cite{Ober}. Another way is to use a transformation procedure, like Fourier or Mellin, to determine the measure. \\
\ni In the Fourier representation, ${\cal W}_m(x; a_r)$ is given by \cite{Aleixo},
 \bea  \label{eq38}
{\cal W}_m(x ; a_r) =\frac{1} {2\pi}\int_{-\infty}^{+\infty}dt \Phi_m(t;a_r)e^{-i x t} ,\quad 
\textrm{where } \quad 
\phi_m(t;a_r) = \sum_{n =0}^{\infty} |K_n^m(a_r)|^2 \frac{(it)^n}{n!}. 
\ena
The explicit expression of ${\cal W}(x, m)$
depends on the term of  $K_n^m(a_r)$ and can be
worked out following standard Handbooks of tabulated integrals
\cite{Marichev}-\cite{Ober}.\\
\ni To use Mellin transformation, we have to rewrite (\ref{eq37}) as 
\bea \label{eq39}
 \int_0^\infty  dx\  x^{n+m}\ g_m(x ; a_r) =  |K_n^m(a_r)|^2, \quad \textrm{where} \quad 
g_m(x; a_r) = \pi {{\cal N}_m^2(x; a_r)} x^{-m} \ \omega_m(x; a_r)\pt
\ena
Let us consider the Meijer's G-function and the Mellin inversion theorem \cite{Mathai} 
{\small
\bea \label{eq40}
  \int_0^\infty dx \, x^{s-1} G_{p,q}^{m,n} 
\left(\alpha x\lv \ba{c} 
a_1,... , a_n  ; a_{n+1},  ..., a_p \\
b_1,... , b_m  ; b_{m+1},  ..., b_q 
\ea \right.\right)
={1 \over \alpha^s}\ {\dis \prod_{j = 1}^m \Gamma(b_j+s) \prod_{j = 1}^n \Gamma(1 - a_j - s) \over  
\dis \prod_{j = m +1}^q \Gamma(1 - b_j -s) \prod_{j = n + 1}^p \Gamma(a_j + s)} \pt
\ena
}
Performing the variable change  $n + m \to s -1, $ \  Eq. (\ref{eq39}) becomes:
\bea \label{eq41}
\int_0^\infty dx x^{s-1} g_m(x;a_r) =  |K_s^m(a_r)|^2 \pt
\ena
In the different examples  of  the next section, $ |K_s^m(a_r)|^2$  in the above relation can be expressed 
in terms of Gamma functions as in the second member of the Mellin inversion theorem (\ref{eq40}). Then comparing the  equations  (\ref{eq40}) and (\ref{eq41}), $ g_m(x;a_r)$ can be identified as the Meijer's G- function:
\bea \label{eq42}
g_m(x;a_r) =  G_{p,q}^{m,n} 
\left(\alpha x\lv \ba{c}  a_1,... , a_n  ; a_{n+1},  ..., a_p \\ b_1,... , b_m  ; b_{m+1},  ..., b_q  \ea \right.\right)\pt
\ena 
  Therefore, the measure $\omega_m$ can be derived from Eq. (\ref{eq39}). 
  {
The overcompleteness of the PA-SIPCS on $\mathfrak{H}_m$ leads to discuss the relation with 
the reproducing kernels.}

 \subsection{Reproducing kernel} 
\ni Define the quantity $\mathcal K(z, z'):={}_{m}\li z';a_r\right.\lv z; a_r\rs_m.$ From 
\begin{eqnarray} \label{kernel0}
 {}_{_{m}}\li z';a_r\right.\lv z; a_r\rs_m 
 & = & \frac{{\cal N}_{m}(|z'|^2; a_r) {\cal N}_m(|z|^2; a_r)}
 { {\cal N}^2_m({z'}^\star z; a_r)}
\end{eqnarray}
we obtain 
\bea\label{kernel00}
 \overline{{}_{_{m}}\li z';a_r\right.\lv z; a_r\rs_m}
 = 
\frac{{\cal N}_{m}(|z'|^2; a_r) {\cal N}_m(|z|^2; a_r)}
 { {\cal N}^2_m(z^\star{z'}; a_r)}
:=\mathcal K(z', z).
\ena
$\mathcal K(z, z')$ is a reproducing kernel through the following result:
  \beprop  The following properties 
\bea
&&(i) \quad \textrm{Hermiticity} \quad \mathcal K(z, z') =\overline{\mathcal K(z', z)}, \\
&&(ii) \quad  \textrm{Positivity}\quad  \mathcal \mathcal K(z, z) >  0,\\
&&(ii)\quad  \textrm{ Idempotence}\quad  \int_{\IC} \ d^2z''\   \omega_m(|z''|^2; a_r) \mathcal K(z, z'')\mathcal K(z'', z') 
\ena  
  are satisfied by the function $\mathcal K$ on $\mathfrak H_m.$
\enprop
  {\bf Proof.} 
\begin{itemize}
\item[(i)] Hermiticity: using  (\ref{kernel0}) and (\ref{kernel00}), we get 
 \bea
\mathcal K(z, z') = {\mathcal K(z', z)}^{\star}.
\ena
\item[(ii)] Positivity: from  (\ref{kernel00}), we obtain
\begin{eqnarray}
 \mathcal K(z, z)= {}_{_{m}}\li z;a_r\right.\lv z; a_r\rs_m  
& = & \frac{{\cal N}_{m}(|z|^2; a_r) {\cal N}_m(|z|^2; a_r)}
 { {\cal N}^2_m(|z|^2; a_r)} = 1  > 0.
\end{eqnarray}
\item[(iii)]Idempotence: let ${\cal I} = \dis \int_{\IC} \ d^2z''\   \omega_m(|z''|^2; a_r) \mathcal K(z, z'')\mathcal K(z'', z')$. Then,
Setting $\xi_{m,m'}(z,z'; a_r) = {\cal N}_m(|z|^2; a_r){\cal N}_m(|z'|^2; a_r)$ gives
{\small
\begin{eqnarray}
{\cal I} & = & \xi_{m,m'}(z,z'; a_r)
\int_{\IC} \ d^2z''\   \omega_m(|z''|^2; a_r)
\frac{{\cal N}^2_m(|z''|^2; a_r)}
{{\cal N}^2_m(zz''^{\star}; a_r) {\cal N}^2_m(z''z'^{\star}; a_r)}\cr
 &=&\xi_{m,m'}(z,z'; a_r)
\sum_{k,p=0}^{\infty}\int_{0}^{\infty}\int_{0}^{2\pi}\frac{e^{-i(k-p)\theta''} \, r''^{k+p}}{| K^m_k(a_r)|^2}
\frac{z^k(z'^{\star})^p}{| K^m_p(a_r)|^2}
{\cal N}^2_m(|z''|^2; a_r) r'' dr''\, d\theta''\omega_m(|z''|^2; a_r)\cr
&=&\xi_{m,m'}(z,z'; a_r)
\sum_{k=0}^{\infty}\frac{z^k(z'^{\star})^k}{| K^m_k(a_r)|^2}
\int_{0}^{\infty}
2\pi \times \frac{dx''}{2}\frac{x''^k}{| K^m_k(a_r)|^2}
\omega_m(x''; a_r){\cal N}^2_m(|z''|^2; a_r)\cr
 &=&\xi_{m,m'}(z,z'; a_r)
\sum_{k=0}^{\infty}\frac{(\sqrt{zz'^{\star}})^{2k}}{| K^m_k(a_r)|^2}
\left\{\int_{0}^{\infty}\frac{x''^{k+m}}{ K^m_k(a_r)|^2} \, g_m(x''; a_r) dx''\right\}\cr
 &=&\frac{{\cal N}_m(|z|^2; a_r){\cal N}_m(|z'|^2; a_r)}{{\cal N}^2_m(zz'^{\star}; a_r)} 
 = \mathcal K(z, z') 
\end{eqnarray}}
which completes the proof.
\end{itemize}
$\hfill{\square}$

\subsection{Photon number statistics}
\ni Here, we deal with  some nonclassical properties, which  will be checked for the constructed PA-SIPCS,   such as 
the photon number distribution (PND), 
 the Mandel Q-parameter and 
the second order correlation function. \\\\
\ni{\it  (i) The PND} \\\\
The probability of finding the vector $|\Psi_n\rangle$ in the  states $|z; a_r \rangle_m$, i.e., the PND which  exhibits  oscillations,  corresponding  to   the probability of finding $n$ quanta in the PA-SIPCS, 
is given by \cite{Aga},\cite{Penson99}
\bea\label{PND}
{\cal P}_n^m(x; a_r) := |\langle n |z;a_r\rangle_m|^2 = {\cal N}_m(x; a_r)^2\,{x^{n-m} \over |K_{n-m}^m(a_r)|^2}, \quad x = |z|^2.
\ena
It reduces to a Poisson distribution for the conventional CS, for $m \to 0$.
This distribution exhibits strong oscillations and its variance is less  than that for a Poisson distribution. \\\\
\ni {\it (ii) The Mandel Q-parameter and the second-order correlation function}\\\\ 
 The Mandel Q-parameter
yields the
information about photon statistics of the quantum states. 
It is  defined as  \cite{Mandel95}:
\bea \label{eq43}
Q := {(\Delta N)^2 \over \li N\rs} - 1, \quad \textrm{with} \quad (\Delta N)^2 := \li N^2\rs - \li N \rs^2 
\ena
and also expressed as  the second-order correlation function given by 
\bea \label{g2}
g^2 := {\langle N^2 \rangle - \langle N \rangle \over \langle N \rangle^2}
\ena
where the mean values of the   operator $N:= H - E_0= B_+ B_-$ and its square in the PA-SIPCS, are defined as:
\bea \label{eq44}
\li N \rs := \, _m\li z;a_r\rv N \lv z; a_r \rs_m,  \quad 
\li N^2 \rs :=\,  _m\li z;a_r\rv N^2 \lv z; a_r \rs_m .
\ena
The Mandel Q-parameter (or the second-order correlation function) determines whether  the PA-SIPCS have a  photon number distribution.
This latter is sub-Poissonian (anti-bunching effect) if $ -1 \le Q < 0$ (or $g^2 <  1$), Poissoinian if $Q = 0$ (or $g^2 =  1$), and 
super-Poissonian (bunching effect) if $Q > 0$ (or $g^2 > 1$). \\
One can check that, for a PA-SIPCS (\ref{eq30}), the expectation values (\ref{eq44}) are:
\bea \label{eq46}
 \li N \rs = {\cal N}_m^2(|z|^2 ; a_r) \sum_{n = 0}^\infty E_{n + m}\ {|z|^{2n}\over  |K_n^m(a_r)|^2 } \ , \ 
\li N^2 \rs =  {\cal N}_m^2(|z|^2 ; a_r) \sum_{n = 0}^\infty E_{n + m}^2\ { |z|^{2n} \over |K_n^m(a_r)|^2 }\pt
\ena
In the next section,  the   quantum statistical features will be explicitly computed for concrete expressions of the  coefficient  $K_n^m(a_r).$ 


\section{Some examples}
\ni  In this section, we  construct the PA-SIPCS for different shape-invariant systems using the
 Infeld and Hull \cite{Infeld} factorization method classification. 
 We  consider  the examples of C, D and A types treated  in \cite{Aleixo} so that, if we put $m = 0$ in our
PA-SIPCS  states,  we recover their corresponding ordinary SIPCS.  
\subsection{Type C and D SIP systems}
\ni They are the  simplest   shape invariant systems. 
The superpotentials  in this case are:
\bea \label{eq47}
W_C(x;a_1) & = & {a_1 +\delta \over x} + {\beta \over 2} x\\ \label{eq48}
W_D(x;a_1) & = & \beta x + \delta
\ena
with the shape invariant condition (\ref{eq8})  written now as:
\bea \label{eq49}
\ba{rcll}
V_2(x;a_1)  & = & V_1(x; a_1 - 1) + 2 \beta  & \quad \textrm{for C-type}\\
V_2(x;a_1)  & = &   V_1(x; a_1) + 2 \beta    & \quad \textrm{for D-type,}
\ea
\ena
the relations between the parameters expressed in the form:
\bea \label{eq50}
\ba{lclcc}
a_{n+1} &  = &   a_n - 1 & \qquad & \textrm{for C-type} \\
a_{n+1} &  = &   a_n = a_{n-1} = \ldots = a_1   & \qquad & \textrm{for D-type} 
\ea \quad  \forall n \in \IN.
\ena
The remainders  are the same  for both C and D systems:
\bea \label{eq51}
{\cal R}(a_1) = {\cal R}(a_2) = \ldots = {\cal R}(a_n) = 2 \beta = \gamma.
\ena
 \subsubsection{D-Type shape invariant systems} 
\ni The products in the expansion coefficient (\ref{eq28}) for D-type systems give  
\bea \label{eq52}
 \prod_{k =1}^m\left(\sum_{s = k}^{n+m}{\cal R}(a_s)\right)  =
\gamma^m\ {\Gamma(n+m+1) \over \Gamma(n+1)}, \quad
\prod_{k =m + 1}^{n+m}\left(\sum_{s = k}^{n+m}{\cal R}(a_s)\right) =
\gamma^n n! .
\ena
The constant values of the potential parameters for D-type SIP imply that we should 
have ${\cal Z}_j =  c = constant$.  Then
\bea \label{eq53}
\prod_{k = m}^{n +m- 1}{\cal Z}_{j + k} = c^n \pt
\ena
Inserting Eqs. (\ref{eq52}), (\ref{eq53}) in  the relation (\ref{eq28}), we obtain   the result for D-type systems:
\bea \label{eq54}
K_n^m(a_r) & = & {\gamma^{(n -m)/2} \over  \sqrt{\Gamma(n + m+ 1)}}\ { n! \over c^n}\pt 
\ena
The unnormalized form of the D-type PA-SIPCS can be written as:
\bea \label{eq55}
\lv z;a_r\rs_m = \sum_{n = 0}^\infty \sqrt{\dis {\Gamma(n + m + 1) \over\gamma^{n - m}}}\ {c^n z^n \over n!} \lv \Psi_{n + m}\rs \pt
\ena
From 
\bea \label{eq56}
{1 \over |K_n^m(a_r)|^2}  & = & \gamma^m{\Gamma(m + n + 1)\over \Gamma(n+1)}\
                                \left({c^2 \over \gamma}\right)^n \ {1 \over n!}
\ena
we deduce the normalization constant  as follows:
\bea\label{eq57}
{\cal N}_m(|z|^2;a_r) & = & \left[\gamma^m\ \Gamma(m+1)\ _1F_1\left(m+1;1;{|cz|^2\over \gamma}\right)\right]^{-1/2}
\ena
  where $_1F_1$ is the generalized  hypergeometric function. It can be obtained  in a more explicit way in terms 
of Meijer's G-function  by:
\bea \label{eq58}
{\cal N}_m(|z|^2;a_r) & = & \left[\gamma^m G_{1,2}^{1,1}\left(\left.- {|cz|^2 \over \gamma} \rv \ba{ccc}-m & ; & \\ 0 & ; & 0 \ea    \right)  \right]^{-1/2}
\ena
where we use the following relation between the generalized  hypergeometric function and the Meijer's G-function \cite{Mathai}:
\bea \label{eq59}
 _pF_q(a_1,\ldots,a_p; b_1,\ldots,b_q; x) = \dis\frac{\dis\prod_{j = 1}^q \Gamma(b_j)}{\dis\prod_{j = 1}^p \Gamma(a_j)}\,
G_{p,q+1}^{1,p}
\left(-x \lv  \ba{rcl} (1 -a_p) & ; & \\ 0 & ; & (1 - b_q) \ea  \right.\right) \pt
\ena
  The explicit form of the PA-SIPCS, defined for any finite $|z|$, can be  read as:
\bea \label{eq60}
 \lv z,a_r\rs_m = {1\over \sqrt{\dis \gamma^m\ \Gamma(m+1)\ _1F_1\left(m+1;1;{|c z|^2\over \gamma}\right)}} \, 
\sum_{n = 0}^\infty \sqrt{\dis {\Gamma(n + m + 1) \over\gamma^{n - m}}}\ {c^n z^n \over n!} \lv \Psi_{n + m}\rs \pt
\ena
For $m = 0$, the expressions of $K_n^m(a_r)$ and ${\cal N}_m(|z|^2;a_r)$ reduce 
 to $h_n(|z|^2;a_r)$ and ${\cal N}(|z|^2;a_r)$, respectively, obtained in \cite{Aleixo}
 for the corresponding ordinary  SIPCS: 
\bea \label{eq61}
\lv z; a_r\rs = \exp\left({-\half {|c z|^2 \over \gamma}}\right) 
\sum_{n = 0}^\infty {c^n z ^n \over \sqrt{\gamma^n n!}}\lv\Psi_n\rs\pt
\ena
Using (\ref{eq33}), the inner product  ${\cal P} = \,  _{m'} \li z';a_r\right.\lv z; a_r\rs_m$ of two different PA-SIPCS 
$\lv z;a_r\rs_m$ and $\lv z';a_r\rs_{m'}$ is given by: 
\bea \label{eq62}
 {\cal P} & = &  {\cal N}_{m'}(|z'|^2; a_r) {\cal N}_m(|z|^2; a_r) \, 
\sum_{n = 0}^\infty {\Gamma(n+m+1) c^{2n} c^{m - m'} ({z'}^\star z)^n {{z'}^\star}^{m - m'} \over 
\gamma^{(n + m -2m')/2} \gamma^{(n -m)/2} (n + m - m')! n!}\pt
\ena
This relation can be expressed in terms of   generalized hypergeometric function $_1F_1$, as:
\bea \label{eq63}
 {\cal P} = \xi(m,m',|z|,|z'|)\ _1F_1\left(m+1; m-m'+1; {|c|^2 {z'}^\star  z \over \gamma}\right)
\ena
where 
$ \xi(m,m',|z|,|z'|) =   {\cal N}_{m'}(|z'|^2; a_r) {\cal N}_m(|z|^2; a_r) c^{m - m'}\ {({z'}^\star)}^{m - {m'}} \gamma^{m'}
\ \dis {\Gamma(m +1)\over \Gamma(m - m'+1)}$.\\ 
 In terms of Meijer's G- function, we have: 
\bea \label{eq64}
 {\cal P} = \chi(m,m',|z|,|z'|)\ G_{2,1}^{1,1} 
\left(\left. - {{z'}^\star z |c|^2 \over \gamma}\rv \ba{ccl}-m & ; & 0 \\ 0 & ; & m' - m \ea \right)
\ena
where 
$ \chi(m,m',|z|,|z'|) =   {\cal N}_{m'}(|z'|^2; a_r) {\cal N}_m(|z|^2; a_r) c^{m - m'}\ {({z'}^\star)}^{m - {m'}} \gamma^{m'}$. 
For $m = m' = 0$,  we recover  the inner product  ${\cal P} = \exp\left[-{|c|^2 \over 2 \gamma}(|z'|^2 + |z|^2 - 2 z'^\star z) \right]$
obtained in  \cite{Aleixo} for the  corresponding SIPCS.\\
\begin{figure}[htbp]
\begin{center}
\includegraphics[width=9cm]{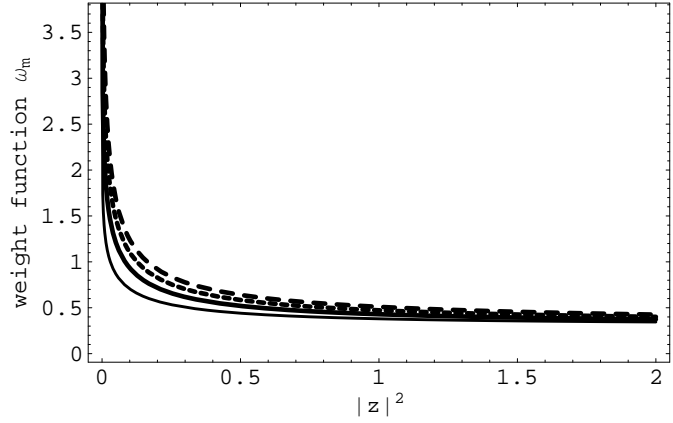}
\end{center}
\ni \caption[]{
Plots of the  weight function (\ref{eq68}) of  the
 PA-SIPCS (\ref{eq60}) versus  $|z|^2$ with $|c|^2 = \gamma,$  for different  values of 
 the photon added number $m$  with 
 $m = 1$ (thin solid  line), $m = 2$ (solid line), $m = 3$ (dot line), and $m = 4$ (dashed line). }
\end{figure}
\ni Let us turn now to the problem of overcompleteness. The relation (\ref{eq39}) gives in this D-type  case:
\bea \label{eq65}
\int_0^\infty dx\ x^{n + m}\ g_m(x; a_r) & = & |K_n^m(a_r)|^2 = 
{\gamma^n n! \ \Gamma(n+1) \over |c|^{2n}\ \gamma^m \ \Gamma(n+m+1)}
\ena
which, after performing the variable change $n + m \to s - 1$, leads to: 
\bea \label{eq66}
\int_0^\infty dx \ x^{s-1}\ h_m(x; a_r) & = & \left(\dis {|c|^2\over \gamma}\right)^{-s} {(\Gamma(s-m))^2 \over \Gamma(s)} 
\ena
where $h_m(x; a_r) = g_m(x;a_r)\gamma^{2m +1}\ c^{-2(m+1)}$. \ 
From the formula (\ref{eq40}) of  the Mellin-inversion theorem, we get:
\bea \label{eq67}
h_m(x; a_r) & = & G_{1,2}^{2,0}\left(\left.{|c|^2\over \gamma} x\rv  \ba{rcl} & ;& 0 \\-m, -m & ; & \ea\right)
\ena
{and the weight function gives
\bea \label{eq68}
 \omega_m(|z|^2;a_r) = {1 \over \pi}{|c|^{2}  \over \gamma}\,
G_{1,2}^{1,1}\left(\left.- {|cz|^2 \over \gamma} \rv \ba{ccc}-m & ; & \\ 0 & ; & 0 \ea    \right) 
 G_{1,2}^{2,0}\left(\left.{|c|^2\over \gamma} |z|^2\rv  \ba{rcl} & ;& m \\0, 0 & ; & \ea\right)
\ena
where we use (\ref{eq58}) and the multiplication formula of the Meijer's G-function \cite{Mathai}
\bea \label{eq68b}
x^\alpha  G_{m,n}^{p,q}\left(x\lv\ba{c} (a_p) \\ (b_q)\ea \right. \right) = G_{m,n}^{p,q}\left(x\lv\ba{c} (a_p + \alpha) \\ (b_q + \alpha)\ea \right. \right).
\ena
Since the measure in equation (\ref{eq36}) must be necessary positive, the
function $\omega_m(|z|^2;a_r)$ must be  a positive function. This is confirmed in FIG. 1,  where  we  represent the weight functions  (\ref{eq68}) for $m = 1, 2, 3, 4$.  We can see that the weight function has a singularity at $x = |z|^2 = 0$ and tends to zero for $x \to \infty$.
}
For $m = 0$, we recover the  measure  $\omega_0(|z|^2; a_r) = \dis{1\over \pi} {|c|^2 \over \gamma}$ obtained in \cite{Aleixo}. Let us note that, by the variable change
 $z\to c  z/\sqrt{\gamma}$, we obtain, for $m = 0$,  the bosonic  coherent states of the  harmonic oscillator     \cite{Pere}. Then, the PA-SIPCS (\ref{eq60}) for $|c|^2 = \gamma$ can be considered as the photon-added coherent states of the harmonic oscillator.\\ 
\quad\newline
\begin{figure}[htbp]
\begin{center}
\begin{minipage}{.45\textwidth}
\includegraphics[width=8.0cm]{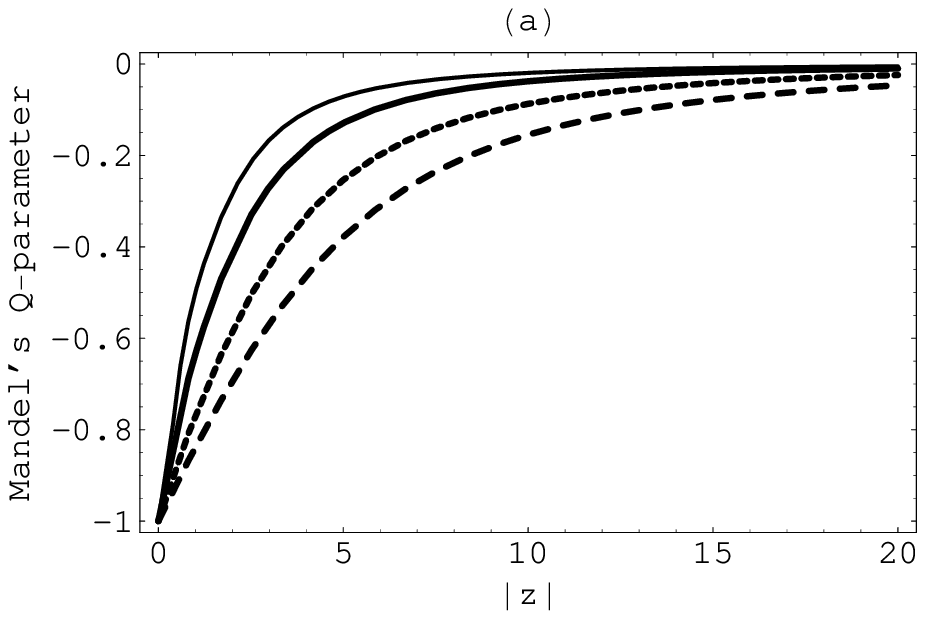}
\end{minipage} 
 \begin{minipage}{.45\textwidth}
\includegraphics[width=8.0cm]{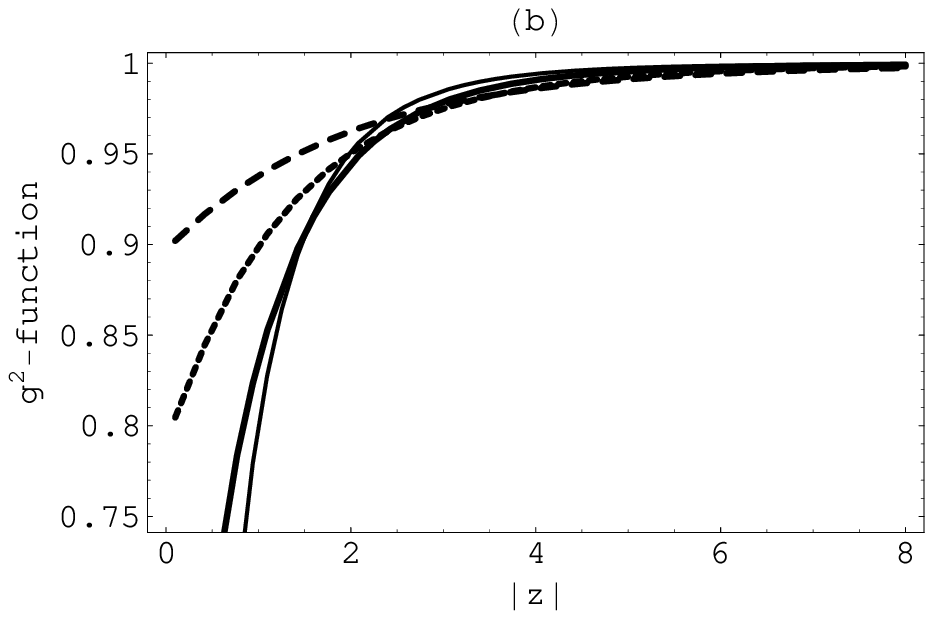}
\end{minipage}  
\end{center} 
\ni \caption[]{ 
Plots of the Mandel Q-parameter  (\ref{QM_Dtype})(a) and the second-order correlation function (\ref{g2_Dtype})(b) of the
 PA-SIPCS (\ref{eq60}) versus  $|z|,$ for different  values of 
 the photon added number $m,$  with 
 $m = 1$ (thin solid  line), $m = 2$ (solid line), $m = 5$ (dot line), and $m = 10$ (dashed line). }
\end{figure}
\newline
\ni To finish with the D-type SI system, let us analyse  the statistical properties of its PA-SIPCS states.  Taking into account the expressions  (\ref{eq54}) and (\ref{eq57}) of the factors $K_n^m(a_r)$ and ${\cal N}_m(|z|^2; a_r)$, the expectation values  $\li N \rs$ and $\li N^2 \rs$ (\ref{eq46})  are given by:
\bea \label{eq69}
\li N \rs & = & m \gamma\  {_2F_2(m+1, m+1 ; 1, m ; |cz|^2/\gamma) \over _1F_1(m+1 ; 1; |cz|^2/\gamma)} \\\label{eq70}
 & & \cr 
\li N^2 \rs & = & m^2 \gamma^2 \ {_3F_3(m+1, m+1, m+1; 1, m, m ;  |cz|^2/\gamma) \over _1F_1(m+1 ; 1; | |cz|^2/\gamma)}.
\ena
Then, the Mandel Q-parameter and the second-order correlation function  can be deduced, respectively,  as:
\bea \label{QM_Dtype}
Q = m \gamma \left( {_3{\cal F}_3(|z|^2; m, \gamma)\over _2{\cal F}_2 (|z|^2; m, \gamma)} -
     {_2{\cal F}_2(|z|^2; m, \gamma)\over _1{\cal F}_1 (|z|^2; m, \gamma)} \right) - 1,
\ena
\bea  \label{g2_Dtype}
g^2 =
{m \gamma\  _3{\cal F}_3(|z|^2; m, \gamma) - _2{\cal F}_2(|z|^2; m, \gamma)
 \over m \gamma\  _2{\cal F}_2(|z|^2; m, \gamma)}\, 
{_1{\cal F}_1(|z|^2; m, \gamma)\over _2{\cal F}_2(|z|^2; m, \gamma)},
\ena
where $_1{\cal F}_1$, $_2{\cal F}_2$ and  $_3{\cal F}_3$ are the generalized hypergeometric functions:
\beano
_1{\cal F}_1 (|z|^2; m, \gamma) & = & _1F_1(m+1 ; 1;  |cz|^2/\gamma) \\
_2{\cal F}_2 (|z|^2; m, \gamma) & = &  _2F_2(m+1, m+1 ; 1, m ;  |cz|^2/\gamma) \\
_3{\cal F}_3(|z|^2; m, \gamma) & = & _3F_3(m+1, m+1, m+1; 1, m, m ;  |cz|^2/\gamma) \pt
\enano
\newline
The PND (\ref{PND}) reads as 
\bea\label{PND_Dtype}
 {\cal P}_n^{(m)}(|z|^2; \gamma) = {\Gamma(n+1) \over \Gamma(m+1)\, _1F_1(1+m ; 1; |cz|^2/\gamma)}\, 
{(|cz|^2/\gamma)^{n-m} \over ((n-m)!)^2}
\ena
\newline
\begin{figure}[htbp]
\begin{center}
\begin{minipage}{.45\textwidth}
\includegraphics[width=7cm]{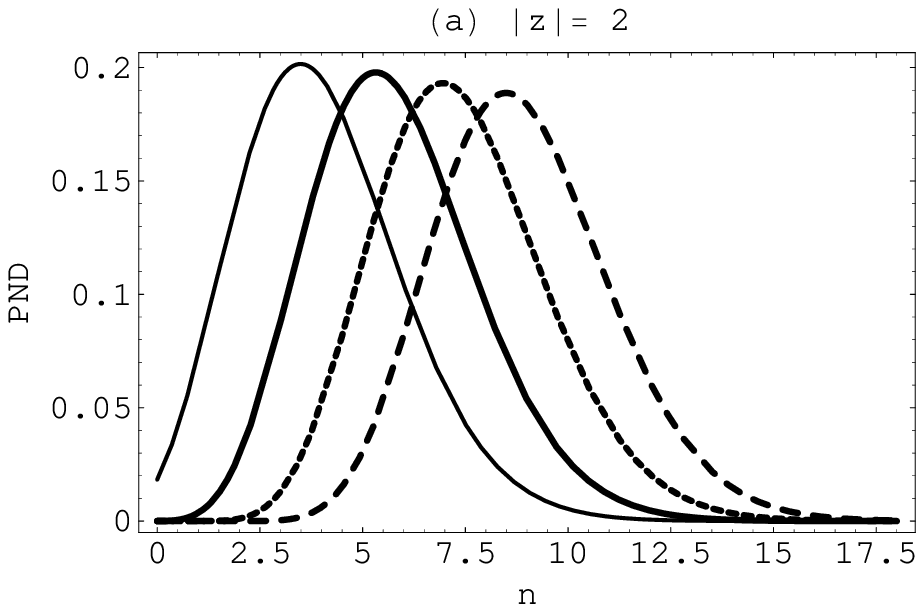}
\end{minipage} 
 \begin{minipage}{.45\textwidth}
\includegraphics[width=7cm]{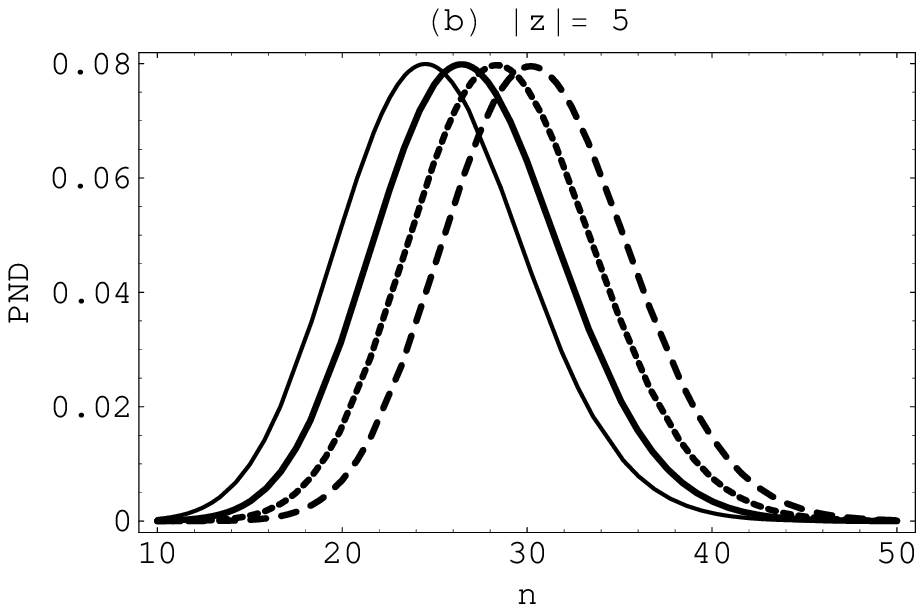}
\end{minipage}  
\end{center} 
\ni \caption[]{
Plots of the photon number distribution  (\ref{PND_Dtype}) of the
  PA-SIPCS (\ref{eq60})  versus  the photon number $n$  for different values of the
photon-added number $m$  for $|z| = 2$(a) and  $|z| = 5$(b),  respectively, with 
$m= 0$ (thin solid  line), $m = 1$ (solid line), $m = 2$ (dot line) and $m = 3$ (dashed line). }
\end{figure}
\newline
In FIG. 2,   the Mandel Q-parameter 
(\ref{QM_Dtype}) and the second-order correlation function (\ref{g2_Dtype}) derived  for the PA-SIPCS (\ref{eq60})
have been plotted in terms of the amplitude $|z|$, in (a) and (b), respectively, 
for different values of the photon-added number $m$,     where $|c|^2 = \gamma$. 
As shown in FIG. 2,   the number $m$ increases as  the Mandel Q-parameter decreases, while 
the second-order correlation function increases.
In addition, the
Mandel Q-parameter increasing,  takes  negative values  for given $|z|$ and $m$ and asymptotically  tends  
to $0$. As for the  second-order correlation function, it  increases
and asymptotically  tends to $1$.    This indicates that  the  PA-SIPCS obey        
     sub-Poissonian distribution statistics. Hence, 
for larger $z,$  the PA-SIPCS (\ref{eq60}) behave like  the states exhibiting Poissonian statistics.\\
In  FIG. 3,  the PND of the   PA-SIPCS (\ref{eq60}), 
 as a function of $|z|$,  is depicted in  (a) and (b)  for   $|z| = 2$ and $|z| = 5, $ respectively,  
 with   different values of the photon-added  number $m$.  As $|z|$  and $m$ increase, the 
 peaks decrease and  shift to the right.

 \subsubsection{C-Type shape invariant systems}
\ni  For C-type systems, we  consider the auxiliary function \cite{Aleixo}: 
\bea \label{eq72}
g(a_j ; c, d) & = & c a_j + d
\ena
where $c$ and $d$ are constants. Using the potential parameter relations  (\ref{eq50}), we can obtain, after a  straightforward computation,  that:
\bea \label{eq73}
\prod_{k = m}^{n + m - 1} g(a_j; c, d) = (-c)^n {\Gamma\left(n + m -a_1 + j-1 -{d\over c}\right) \over 
\Gamma\left(m - a_1 + j-1 -{d\over c}\right)}\pt
\ena
 Defining the functional
${\cal Z}_j := \sqrt{g(a_1; -\gamma, 1)} \, e^{- i \alpha {\cal R}(a_1)}$, 
we get:
\bea \label{eq74}
\prod_{k = m}^{n + m -1} Z_{j + k} = \sqrt{\gamma^n \Gamma( - \rho + m + n) \over 
\Gamma(  - \rho + m)}\, e^{- i\alpha n \gamma}
\ena
where we set $\rho = a_1 - {1\over \gamma}$. 
This last relation, with (\ref{eq52})  inserted  in (\ref{eq28}), gives 
for C-Type systems:
\bea \label{eq75}
K_n^m(a_r) & = & \left[{ \Gamma(-\rho + m ) \Gamma(n+1)^2 \over   \gamma^{ m} \Gamma(-\rho + m + n) \Gamma(n + m +1 )
 } \right]^{1/2}\ 
e^{i \alpha n \gamma}\pt
\ena
The normalized form of the C-type PA-SIPCS, defined on the unit open disc $|z|< 1$,  can be written as:
\bea \label{eq76}
 \lv z;a_r\rs_m =  {\cal N}_m(|z|^2;a_r)\sum_{n = 0}^\infty 
\sqrt{\dis { \gamma^{m} \Gamma(-\rho + m + n) \Gamma(n + m +1 ) \over \Gamma(-\rho + m )} }\,  
e^{-i \alpha n \gamma}\ 
 {z^n \over n!} \lv \Psi_{n + m}\rs 
\ena
 where the normalization factor  is given by: 
\bea  \label{eq77}
{\cal N}_m(|z|^2;a_r) & = & \left[\gamma^m\ \Gamma(m+1)\ _2F_1\left(m- \rho, m +1 ;1;|z|^2\right)\right]^{-1/2}.
\ena
Using Eq. (\ref{eq59}), the normalization factor can also be formulated  in terms of Meijer's G-function as:  
\bea  \label{eq78}
{\cal N}_m(|z|^2;a_r) & = & \left[{\gamma^m \over \Gamma(m - \rho)} \ 
G_{2,2}^{1,2}\left(-|z|^2 \lv \ba{ccc}-m, 1-m + \rho & ; & \\ 0 & ; & 0 \ea \right.    \right)  \right]^{-1/2}\pt
\ena
For $m = 0$, $K_n^m(a_r)$ and ${\cal N}_m(|z|^2;a_r)$ reduce to the equivalent quantities $h_n(a_r)$ and ${\cal N}
(|z|^2;a_r)$, respectively,  for the corresponding ordinary SIPCS \cite{Aleixo} :
\bea \label{eq79}
K_n^0(a_r) & = &  \sqrt{\Gamma(-\rho) \Gamma(n + 1) \over \Gamma(n - \rho)}\, e^{i \alpha n \gamma} = h_n(a_r) \\ \label{eq80}
{\cal N}_0(|z|^2; a_r) & = & (1 - |z|^2)^{-\rho/2} = {\cal N}(|z|^2; a_r)\pt
\ena
Then, the PA-SIPCS correspond to  ordinary SIPCS obtained in \cite{Aleixo} :
\bea \label{eq81}
\lv z;a_r\rs =  (1 - |z|^2)^{-\rho/2} \sum_{n = 0}^\infty \sqrt{\Gamma(n - \rho) \over \Gamma(-\rho) \Gamma(n + 1)}\, e^{-i \alpha n \gamma}\,
z^n \lv \Psi_n\rs
\ena
which are the Perelomov CS for the SU(1,1) group \cite{Perelomov}. Consequently, Eqs. (\ref{eq76}) and (\ref{eq77}) can be considered as the PA-SIPCS  associated to the Perelomov CS. \\
The inner product of two different PA-SIPCS $\lv z; a_r\rs_m$ and $\lv z';a_r\rs_{m'}$ follows from Eq (\ref{eq33}):
\bea \label{eq82}
 _{m'}\li z';a_r\rv \left. z; a_r\rs_m = F(\rho, m, m')\ 
_2F_1(m  - \rho, m +1 ; m - m'+ 1;{ z'}^\star z)
\ena
where 
{\small \bea \label{eq83}
 F(\rho, m, m') = {\cal N}_{m'}(|z'|^2;a_r)\ {\cal N}_m(|z|^2;a_r) \ {\Gamma(m +1) \over \Gamma(m -m'+1)}\ 
\sqrt{ \gamma^{(m + m')}\Gamma( - \rho + m) \over ¯\Gamma(  - \rho + m')}
\ e^{i \alpha (m - m')\gamma}.
\ena }
For $ m = m' = 0$, we recover the result for the   SIPCS obtained in  \cite{Aleixo}:
{\small
\bea \label{eq84}
  _{0}\li z';a_r\rv \left. z; a_r\rs_0 & = & {\cal N}(|z'|^2; a_r) {\cal N}(|z|^2; a_r)_2F_1(-\rho, 1; 1; {z'^\star} z) 
                                   = \left[{\sqrt{(1 - |z|^2) (1 - |z'|^2)}\over 1 - {z'^\star} z } \right]^{-\rho}\pt
\ena}
\begin{figure}[htbp]
\begin{center}
\includegraphics[width=9cm]{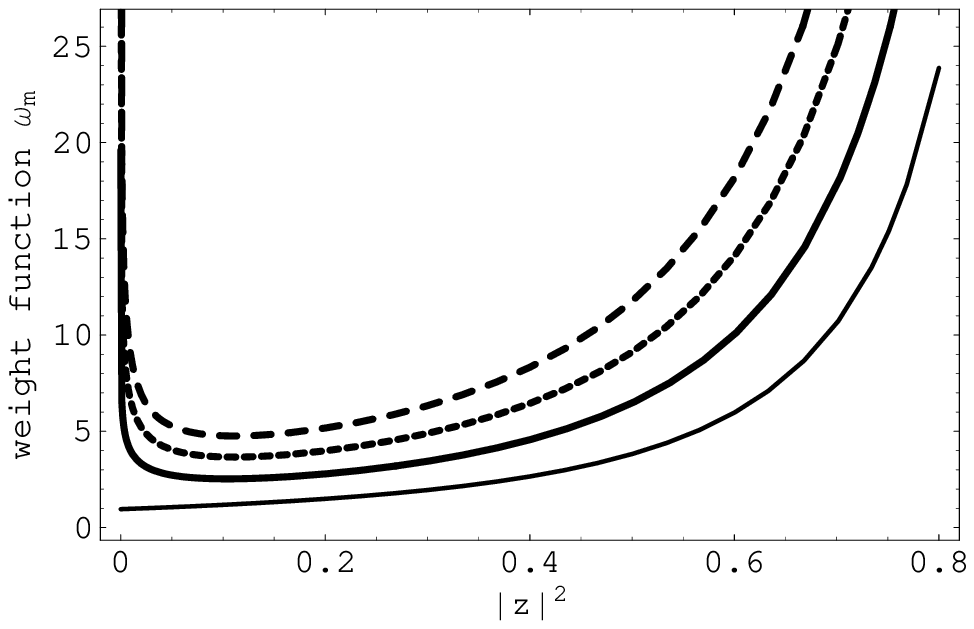}
\end{center}
\ni \caption[]{ 
Plots of the  weight function (\ref{eq88}) of  the
 PA-SIPCS (\ref{eq76}) versus  $|z|^2$ for different  values of 
 the photon added number $m$  with 
 $m = 0$ (thin solid  line), $m = 1$ (solid line), $m = 2$ (dot line), and $m = 3$ (dashed line). }
\end{figure}
\ni Taking into account the expression (\ref{eq75}) of the expansion coefficient, the overcompleteness relation (\ref{eq39}) becomes 
\bea \label{eq85}
 \int_0^\infty dx \ x^{n +m}g_m(x;a_r) = {\Gamma(- \rho + m) \Gamma(n +1) \Gamma(n +1) 
 \over  \gamma^m \Gamma(- \rho + m + n) \Gamma(n + m +1)}\pt
\ena
Making the variable change $n + m \to s -1$, we get
\bea \label{eq86}
 \int_0^\infty dx \ x^{s - 1}\ h_m(x;a_r) = { \Gamma( s - m) \Gamma(s - m) 
 \over  \Gamma(s) \Gamma( s  - \rho -1)}
\ena
where $h_m = \dis {\gamma^m \over \Gamma(- \rho + m)}\times g_m$. Identifying Eq. (\ref{eq86}) and  taking into account the formuula (\ref{eq40}) given by the Mellin inversion theorem, we deduce 
\bea \label{eq87}
h_m =  G_{2,2}^{2,0}\left(\left. |z|^2\rv  \ba{rcl} & ;& 0 , -\rho -1 \\
-m, -m & ; & \ea\right).
\ena
{Then the measure is derived as 
{\small 
\bea \label{eq88}
 \omega_m(|z|^2; a_r &  = & {1\over \pi}
G_{2,2}^{1,2}\left(-|z|^2 \lv \ba{ccc}-m, 1-m + \rho & ; & \\ 0 & ; & 0 \ea \right. \right)
  G_{2,2}^{2,0}\left(|z|^2\lv  \ba{rcl} & ;& m ,m -\rho -1 \\
0, 0 & ; & \ea \right.\right) 
\ena}
 where we  use (\ref{eq78}) and the multiplication formula of the Meijer's G-function (\ref{eq68b}). The  measure  (\ref{eq88}) is positive for $\rho < -1$ as shown  on the representation in FIG. 4.  
  for $\rho = - 2$ and  for $m = 0, 1, 2, 3$.  We remark that the measure $\omega_m(|z|^2; a_r)$ for  $m \ne 0$ has the same behaviour as  the measure  corresponding to the  conventional coherent states ($m  = 0$). It presents   singularities at $x = 0$ and $x = 1$.}
For $m = 0$, the Meijer's G-function reduces to $\dis {(1 - |z|^2)^{-\rho -2} \over \Gamma(-\rho -1)}$ and  we recover 
the measure  $-{1\over \pi}(1+\rho) (1 -|z|^2)^{-2}$ obtained in \cite{Aleixo}.
\quad \newline
\begin{figure}[htbp]
\begin{center}
\begin{minipage}{.45\textwidth}
\includegraphics[width=8.0cm]{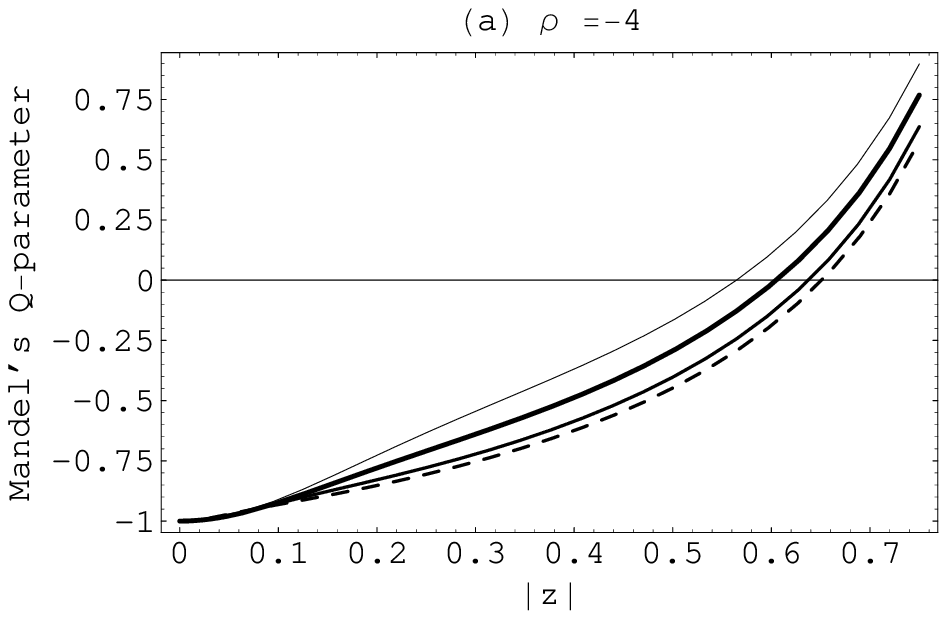}
\end{minipage} 
 \begin{minipage}{.45\textwidth}
\includegraphics[width=8.0cm]{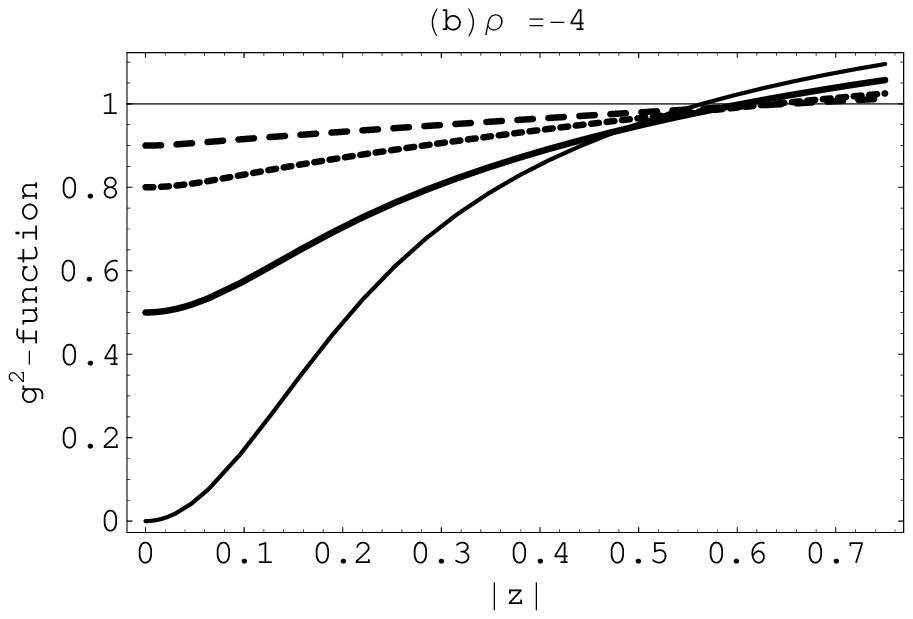}
\end{minipage}  
\end{center} 
\ni \caption[]{
Plots of the Mandel Q-parameter  (\ref{QM_Ctype})(a)  and the second-order correlation function (\ref{g2_Ctype})(b)  of the
 PA-SIPCS (\ref{eq76}) versus  $|z|$  with the  parameter $\rho = - 4$   for different values of 
 the photon added number $m$  with 
 $m = 1$ (thin solid  line), $m = 2$ (solid line), $m = 5$ (dot line), and $m = 10$ (dashed line).}
\end{figure}
\quad \newline
We end the discussion on C-type SI systems by the study of the  statistical properties of their  PA-SIPCS states. 
 Inserting  the expressions 
 (\ref{eq75}) and (\ref{eq77}) of the factors $K_n^m(a_r)$ and ${\cal N}_m(|z|^2; a_r)$
  in Eq. (\ref{eq46}),   
we obtain  the expectation values of $\li N \rs$ and $\li N^2 \rs$  as:
\bea \label{eq89}
 \li N \rs &=  &m \gamma\  {_3F_2(- \rho + m, m+1, m+1 ; 1, m ; |z|^2) \over 
_2F_1(- \rho + m, m+1 ; 1; |z|^2)} \\\label{eq90}
 \li N^2 \rs &= & m^2 \gamma^2 \ {_4F_3( - \rho + m, m+1, m+1, m+1; 1, m, m ;  |z|^2) \over 
_2F_1(- \rho + m, m+1 ; 1; |z|^2)}\pt
\ena 
Then, the Mandel Q-parameter and the second-order correlation function can be deduced, respectively, as:
\bea \label{QM_Ctype}
Q = m \gamma \left( {_4{\cal F}_3(|z|^2; m, \rho)\over _3{\cal F}_2 (|z|^2; m, \rho)} -
     {_3{\cal F}_2(|z|^2; m, \rho)\over _2{\cal F}_1 (|z|^2;  m, \rho)} \right) - 1
\ena
\bea  \label{g2_Ctype}
g^2 =
{m \gamma\  _4{\cal F}_3(|z|^2; m, \rho) - _3{\cal F}_2 (|z|^2; m, \rho)
 \over m \gamma\ _3{\cal F}_2 (|z|^2; m, \rho)}\, 
{_2{\cal F}_1 (|z|^2; m, \rho) \over _3{\cal F}_2 (|z|^2; m, \rho)},
\ena
where $_2{\cal F}_1$, $_3{\cal F}_2$ and  $_4{\cal F}_3$ are the generalized hypergeometric functions:
\beano
_2{\cal F}_1 (|z|^2; m, \rho) & = & _2F_1(- \rho + m, m+1 ; 1; |z|^2)\\
_3{\cal F}_2 (|z|^2; m, \rho) & = &  _3F_2(- \rho + m, m+1, m+1 ; 1, m ; |z|^2) \\
_4{\cal F}_3(|z|^2; m, \rho) & = & _4F_3(- \rho + m, m+1, m+1, m+1; 1, m, m ;  |z|^2) \pt
\enano
\begin{figure}[htbp]
\begin{center}
\begin{minipage}{.45\textwidth}
\includegraphics[width=8cm]{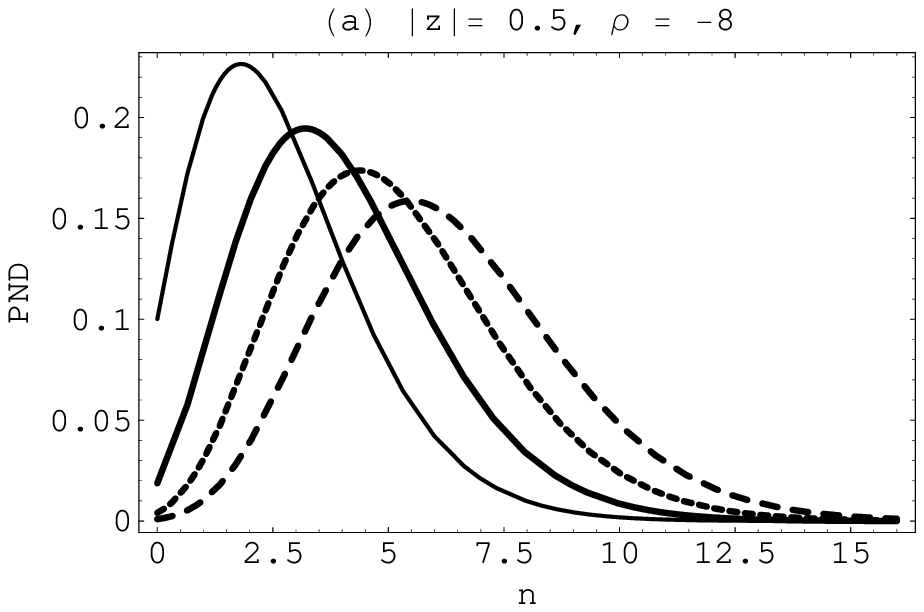}
\end{minipage} 
 \begin{minipage}{.45\textwidth}
\includegraphics[width=8cm]{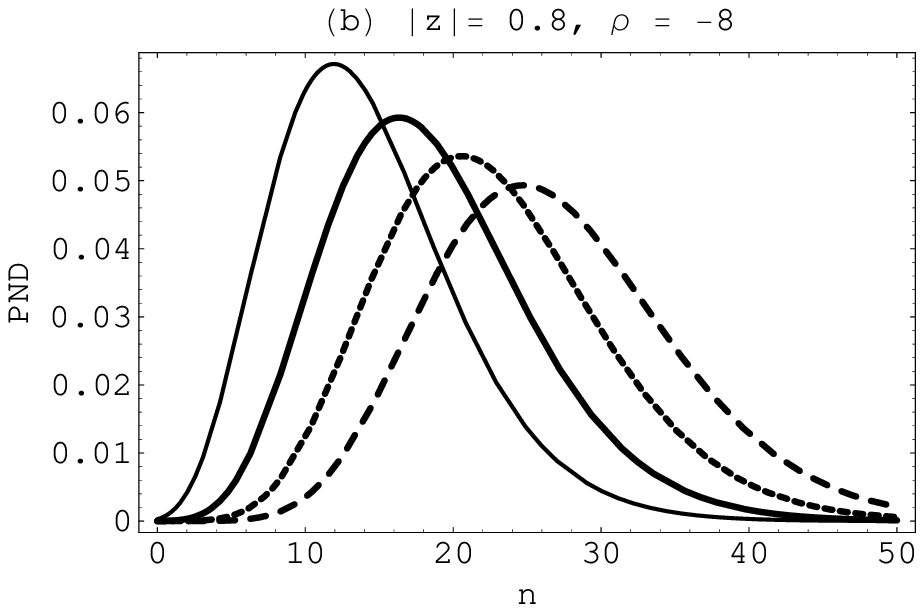}
\end{minipage}  
\end{center} 
\ni \caption[]{ Plots of the photon number dstribution  (\ref{PND_Dtype}) of the
  PA-SIPCS (\ref{eq76}) versus  the photon number $n$  for different values of the
photon-added number $m$ with parameter $\rho  = - 8$   for  $|z| = 0.5$(a) and  $|z| = 0.8$(b), respectively, with 
$m= 0$ (thin solid  line), $m = 1$ (solid line), $m = 2$ (dot line) and $m = 3$ (dashed line). }
\end{figure}
The PND (\ref{PND}) reads as 
\bea\label{PND_Ctype}
 {\cal P}_n^{(m)}(|z|^2;  \rho) = {\Gamma(n+1) \Gamma(n- \rho)\over \Gamma(m- \rho) \Gamma(m+1)\, _2{\cal F}_1 (|z|^2; m, \rho)}\, 
{(|z|)^{2(n-m)} \over ((n-m)!)^2}.
\ena
In FIG. 5,    the Mandel Q-parameter (\ref{QM_Ctype}) and   the second-order correlation 
function (\ref{g2_Ctype}) derived in  the PA-SIPCS (\ref{eq76}) are plotted in terms of the 
amplitude $|z|$,   in (a) and (b), respectively, for different values of the photon-added number $m$
with the potential parameter $\rho = -4$.  
As $|z|$ increases from  $0$
 to $1$, the Mandel Q-parameter increases from negative values to positive ones, while the second-order correlation function increases from $g^2 < 1$ to  $g^2 > 1$.  
\ni Thus, there exists a value $z_0$, depending on the parameter $\rho$, such  that the 
 PASIPCS (\ref{eq76})  exhibit sub-Poisonnian distribution  for $0 < |z| < |z_0|$, Poissonian for $|z| = |z_0| $ and super-Poisonnian  distribution for $|z_0| < |z| < 1$.\\
In FIG. 6,  the PND obtained in  the PA-SIPCS (\ref{eq76})
 as a function   of $|z|$, is depicted in 
 (a)  and (b)  for  $|z| = 0.5$ and $|z| = 0.8$, respectively,  for different values of the photon-added  number $m$. Increasing 
 the number $m$ shifts   the PND.  
 The peaks position increases with  the photon-added number  $m$ and  the amplitude $|z|$, 
while their intensities decrease.

\subsection{Type A and B SIP systems}
\ni The superpotentials for these types of SI systems  are:
\bea \label{eq92}
 W_{A}(x; a_1) & = & \beta(a_1 + \gamma)\cot[u(x)]+ {\delta \over \sin[u(x)]}, \quad u(x)\equiv \beta(x + \lambda) \\ \label{eq93}
 W_{B}(x;a_1) & = & i\beta (a_1 + \gamma) + \delta e^{-i \beta x}
\ena
\ni where $\beta$ is  a real  constant for A-type or  pure imaginary for B-type and  $\gamma, \lambda, \delta$ are real constants. For both cases, the remainder  in the shape invariant condition (\ref{eq8}) is  ${\cal R}(a_1)  = \beta^2[2(a_1 + \gamma) - 1] $, the potential parameters being related as: $a_{n+1} = a_n - 1$.
Since we are interested to bound states Hamiltonian, we restrict our attention to A-type systems. 
For these systems, the products in terms of the quantity ${\cal R}(a_s)$ in the numerator and denominator of the coefficient $K_n^m(a_r)$, see Eq. (\ref{eq28}), can be read, respectively, as:
\bea \label{eq94}
\prod_{k = m+1}^{n + m} \left(\sum_{s=k}^{n+m} {\cal R}(a_s) \right) & = &  \beta^{2n}\frac{\Gamma(n+1) \Gamma(2 n+ 2 m + 2 \rho)}
                                                      {\Gamma(n + 2m + 2\rho)} \\\label{eq95}
\prod_{k =1}^{m}\left(\sum_{s=k}^{n+m} {\cal R}(a_s) \right)& = &  \beta^{2m}\frac{\Gamma(n+m+1) \Gamma(n+ 2 m + 2 \rho)}
                                                      {\Gamma(n+1)\Gamma(n+ m + 2\rho)} 
\ena
where we set $\rho = -(a_1 + \gamma)$. 
The explicit form of the expansion coefficient $K_n^m(a_r)$ depends on the choice of the functional ${\cal Z}_j$.  
We adopt the functionals used in \cite{Aleixo} for commodity of comparison. 
\subsubsection{First choice of the functional ${\cal Z}_j$}
\ni First we make the choice ${\cal Z}_j = c$,  where $c$ is a real constant. 
Then $\dis \prod_{k =m}^{n+m-1}{{\cal Z}_{j+k}} = c^n$. Inserting this and the  results (\ref{eq94}) and (\ref{eq95}) in  (\ref{eq28}), we obtain the expansion coefficient as:
\bea \label{eq96}
K_n^m(a_r) & = & {1\over \kappa^{m}}\ \sqrt{\frac{\Gamma(n+1)^2\ \Gamma(2n+2m+2 \rho)\ \Gamma(n+ m +2 \rho)}
{\Gamma(n+m+1) \Gamma(n+ 2 m + 2 \rho)^2}}                                                   
\ena
where we set  $\kappa = \beta = c$.  \\
\ni {\it (i) Normalization} \\
The normalization factor, in terms of the generalized hypergeometric functions $_3F_4$, can  readily be 
 deduced from the above relation as:
{\footnotesize
\bea \label{eq97}
{\cal N}_m(|z|^2; a_r) &=  & \left[ 
 \xi(m,\rho) {}_3F_4\left(m+1, 2m+2\rho,2m+2\rho ;  1, m+\rho,m+2\rho, m+\rho + 1/2;  {|z|^2\over 4}\right) \right]^{-1/2} 
\ena}
where $\xi(m,\rho) = \dis {\kappa^{2m}\Gamma(m+1)\Gamma(2m + 2 \rho)\over \Gamma(m + 2 \rho)} $. In terms of Meijer's G-function, we have:
{
\bea \label{eq98}
{\cal N}_m(|z|^2; a_r) & = &
\left[ \chi(m,\rho)\ 
 G_{3,5}^{1,3}\left(-{|z|^2\over 4}\left|\ba{l}
                                    -m, 1 - 2 m - 2\rho , 1 - 2 m - 2\rho ;  \\ 
                      0 , 0, 1-m - \rho, 1-m - 2\rho, 1/2 -m - \rho \ea \right.\right) \right]^{-1/2}
\ena}
with $\chi(m,\rho)  = \dis {\kappa^{2m} \Gamma(m +  \rho) \Gamma(m +  \rho + \half) \over \Gamma(2m + 2 \rho)}$ and where we use  the more compact  notation of Meijer's G-function.
\bea \label{eq98b}
 G_{p,q}^{m,n} 
\left( x\lv \ba{c} 
a_1,... , a_p \\
b_1,... , b_q 
\ea \right.\right)\pt
\ena
The explicit form of the PA-SIPCS are:
\bea \label{eq99}
  \lv z; a_r\rs_m = {\cal N}_m(|z|^2; a_r) \kappa^m 
\sum_{n = 0}^{\infty} \sqrt{\frac{\Gamma(n+m+1) \Gamma(n+ 2 m + 2 \rho)^2}
                   {\Gamma(n+ m + 2 \rho)\Gamma(2n+2m+2 \rho)}}\,  {z^{n}\over n!} \lv n+m\rs
\ena
defined on the whole complex  plane. 
  For $m = 0$, we recover the expansion coefficient  and the normalization factor  obtained in \cite{Aleixo} for the generalized SIPCS:
{\footnotesize
\bea \label{eq100}
 K_n^0 & = & \sqrt{\Gamma(n+1) \Gamma(2\rho + 2n) \over \Gamma(2\rho + n)} = h_n(a_r), \cr
 {\cal N}_0(|z|^2; a_r) & = & 
\left[{}_1F_2\left(2\rho; \rho, \rho + 1/2 ; {|z|^2\over 4}\right)\right]^{-1/2} = {\cal N}(|z|^2; a_r)
\pt
\ena}
For giving $m$, fixing $\rho = 1/2$, the states (\ref{eq99}) become 
\bea \label{eq101}
\lv z; a_r\rs_m = {\cal N}_m(|z|^2; a_r) \kappa^m \sum_{n = 0}^\infty \sqrt{\frac{(n+2m)^2!}{(2 n+ 2m)!}}{z^n\over n!}\, 
\lv n + m\rs
\ena
with 
\beano
 {\cal N}_m(|z|^2; a_r) =
\left[\kappa^{2m} \Gamma(2m+1){}_2F_3\left(2m+1,2m+1; 1,m+1, m+1/2; {|z|^2\over 4}\right)\right]^{-1/2}\pt
\enano
Setting $m = 0$ in (\ref{eq101}), we recover  the ordinary SIPCS obtained in \cite{Fukui}: 
\bea\label{eq102}
\lv z;a_r\rs = \sqrt{\sech(|z|)}\,  \sum_{n = 0}^\infty {z^n \over \sqrt{2n!}}\, \lv n\rs\pt
\ena
\ni {\it (ii) Non-orthogonality} \\
The inner product of two different PA-SIPCS $\lv z; a_r\rs_m$ and $\lv z';a_r\rs_{m'}$ follows from Eq (\ref{eq33}):
{\small 
\beano
& &  _{m'}\li z';a_r\rv \left. z; a_r\rs_m  = \chi(z',z, m, m',\rho)\times \\
 & & \times _3F_4\left(m +1,2m+2\rho,m+m'+2\rho ; m - m'+ 1, m+2\rho, m+\rho, m+\rho+\half; 
{{ z'}^\star z\over 4}\right)
\enano
}
where 
{\small \beano
\chi(z',z, m, m',\rho) = {\cal N}_{m'}(|z'|^2;a_r)\ {\cal N}_m(|z|^2;a_r) \ {{ z'}^\star}^{(m-m')} \kappa^{(m+m')}
 {\Gamma(m +1)  \Gamma(m +m'+2\rho) \over \Gamma(m -m'+1)  \Gamma(m +2\rho)}.\\
\enano }
\begin{figure}[htbp]
\begin{center}
\includegraphics[width=9cm]{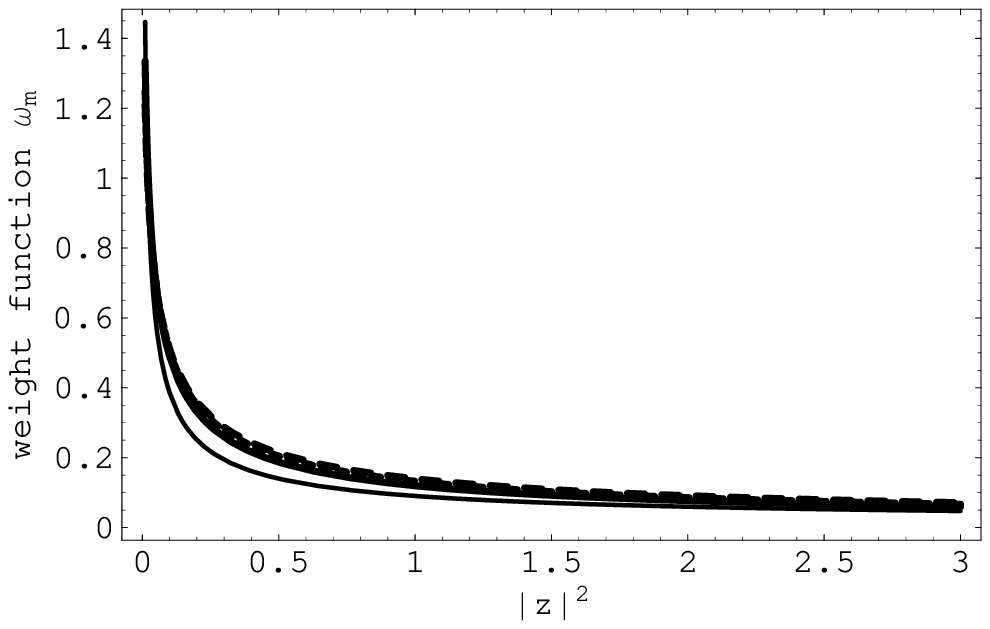}
\end{center}
\ni \caption[]{ 
Plots of the  weight function (\ref{eq105}) of  the
 PA-SIPCS (\ref{eq99}) versus  $|z|^2$ for different  values of 
 the photon added number $m$  with 
 $m = 0$ (thin solid  line), $m = $ (solid line), $m = 3$ (dot line), and $m = 4$ (dashed line). }
\end{figure}
\ni {\it (iii) Overcompleteness} \\
The non-negative weight function $\omega_m(|z|^2;a_r)$ is related to the  function $g_m$  satisfying  (\ref{eq39}):
\bea \label{eq103}
\int_0^\infty dx\,x^{n+m}\, g_m(x;a_r) & = &\xi(x,n,m,\rho)\,\cr
& &  \times {\Gamma(n+1)^2 \Gamma(n + m + 2\rho) \Gamma(n + m + \rho) \Gamma(n + m + \rho +\half) \over 
                                                                     \Gamma(n + m + 1) \Gamma(n + 2m + 2\rho)^2}
\ena
where $x$ stands for $|z|^2$,  $\xi(x, n,m,\rho) = \dis {2^{(2 n + 2 m + 2\rho)}\over 2 \sqrt{\pi} \kappa^{2 m}}  $  and 
$\omega_m = \dis {x^m g_m(x;a_r) \over \pi N_m^2(x; a_r)}$. 
After variable change $n + m \to s -1$, (\ref{eq103}) becomes 
\bea \label{eq104}
\int_0^\infty dx \,x^{s - 1}\, h_m(x;a_r)  = \left({1\over 4}\right)^{-s}\,
                                          {\Gamma(s - m)^2 \Gamma(s - 1 + 2\rho) \Gamma(s - 1 + \rho)\Gamma(s - 1/2 + \rho)
                                          \over \Gamma(s) \Gamma(s - 1 + 2 \rho + m)^2}
\ena
where $h_m(x; a_r)$ is related to $g_m(x;a_r)$ as: $g_m = {1\over 2 \sqrt{\pi}} \kappa^{-2m} 2 ^{2(\rho -1)} h_m(x;a_r)$. Then, using the Mellin inversion theorem in terms of Meijer's G-function (\ref{eq40}), {we deduce:
\bea \label{eq105}
\omega_m(|z|^2;a_r) & = &{1\over 4\pi}
 G_{3,5}^{1,3}\left(-{|z|^2\over 4}\left|\ba{l}
                                    -m, 1 - 2 m - 2\rho , 1 - 2 m - 2\rho ;  \\ 
                      0 , 0, 1-m - \rho, 1-m - 2\rho, 1/2 -m - \rho \ea \right.\right)\times \cr
& &\times 
 G_{3,5}^{5,0}\left({|z|^2\over 4}\left|\ba{l}
 m,-1+2\rho + 2 m,-1+2\rho +2 m \\ 
 0,0,2\rho + m - 1, -1+ m+\rho,-1/2 + m +\rho 
  \ea \right.\right)
\ena
where we  use (\ref{eq98}). The weight function (\ref{eq105}) is positive for  the parameter $\rho > 0$ as confirmed in FIG. 7, where the curves are represented for  $\rho =  \half$ and for $m = 0, 1, 2, 3$. All the functions are positive for $x = |z|^2\in \IR_+$ and tend asymptotically to the measure of the conventional CS $(m = 0)$. The measure has a singularity at $x =  0$  and tends to zero for $x \to \infty.$}
For $m = 0$ and  $\rho = 1/2$ we have  $h_0 = \dis 2\sqrt{\pi}{e^{-|z|} \over |z| }$ and ${\cal N}_0^2(|z|^2;a_r) = \sech(|z|)$. 
Then, we recover  the weight function  $\omega(|z|^2; a_r) = \dis{\cosh(|z|)\over  \pi}{e^{-|z|} \over 2 |z|}$  obtained in \cite{Fukui}. \\
{\it (iv) Statistical properties} 
\begin{figure}[htbp]
\begin{center}
\begin{minipage}{.45\textwidth}
\includegraphics[width=8cm]{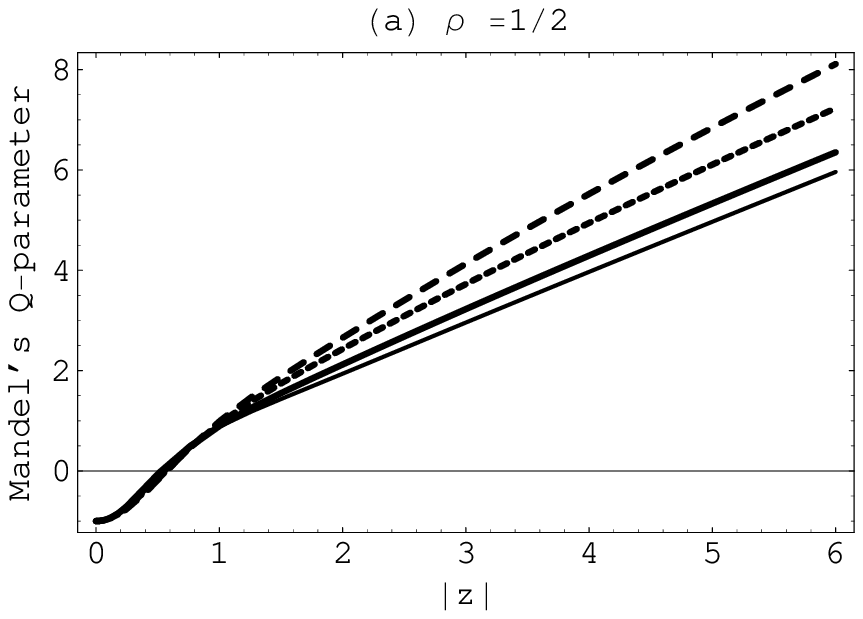}
\end{minipage} 
 \begin{minipage}{.45\textwidth}
\includegraphics[width=8cm]{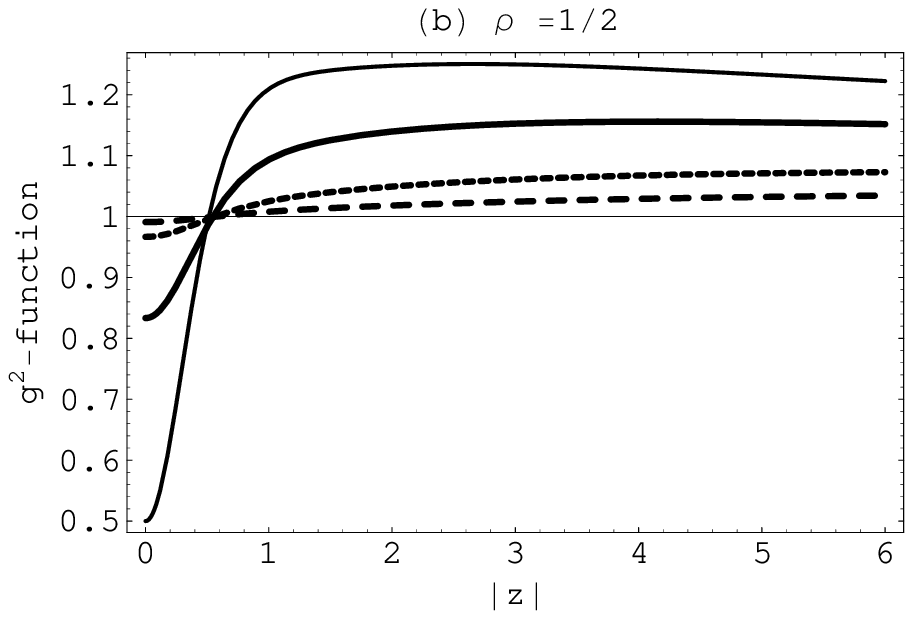}
\end{minipage}  
\end{center} 
\ni \caption[]{
Plots of the  Mandel Q-parameter  (\ref{QM_Atype1})(a)  and the second-order correlation function  (\ref{g2_Atype1})(b)  of the
 PA-SIPCS (\ref{eq99}) versus  $|z|$  with the  parameter $\rho = \half$   for different values of 
 the photon added number $m$  with 
 $m = 1$ (thin solid  line), $m = 2$ (solid line), $m = 5$ (dot line), and $m = 10$ (dashed line).
}
\end{figure}
\newline
\ni We consider now the statistical properties of the state (\ref{eq99}). 
The expectation values $\li H\rs$ and $\li H^2\rs$ result from the expressions (\ref{eq96}) and (\ref{eq97}) of $K_n^m$ and ${\cal N}_m$, 
respectively, as:
{\small
\bea \label{eq107}
\li N\rs  & = &  \kappa^2 m(m+ 2 \rho) {_5{\cal F}_6(|z|^2; m,\rho) \over _3{\cal F}4(|z|^2; m, \rho)}\quad, \quad 
\li N^2\rs =  \kappa^4 m^2(m+ 2 \rho)^2 {_7{\cal F}_8(|z|^2; m,\rho) \over  _3{\cal F}_4(|z|^2 ; m, \rho)}\pt
\ena}
\ni Then, the Mandel Q-parameter and the second-order correlation function are  derived, respectively, as:
\bea \label{QM_Atype1} 
Q &= &\kappa^2 m (m + 2 \rho) \left( {_7{\cal F}_8(|z|^2; m, \rho)\over _5{\cal F}_6 (|z|^2; m, \rho)} -
     { _5{\cal F}_6(|z|^2;  m, \rho)\over _3{\cal F}_4 (|z|^2;  m, \rho)} \right) - 1\\ \label{g2_Atype1}
g^2& =&
{m \kappa^2 (m + 2\rho)\ _7{\cal F}_8(|z|^2; m,\rho) - _5{\cal F}_6(|z|^2; m,\rho)
 \over m \kappa^2 (m + 2\rho)\ _5{\cal F}_6(|z|^2; m,\rho)}\, 
{_3{\cal F}_4(|z|^2; m, \rho) \over _5{\cal F}_6(|z|^2; m,\rho)},
\ena
where $_3{\cal F}_4$, $_5{\cal F}_6$ and  $_7{\cal F}_8$ are the generalized hypergeometric functions:
{\small
\beano
  _3{\cal F}_4 (|z|^2; m, \rho) & = &  _3F_4\left(\ba{lcr} m+1,  2m+2\rho,2m+2\rho & ; &\\ 
                                                      1, m+\rho,  m+2\rho, m +\rho +\half & ; & {|z|^2/ 4}\ea\right) \\
  _5{\cal F}_6 (|z|^2; m, \rho) & = &  _5F_6\left(\ba{lcr} m+1, m+1, 2m+2\rho, 2m+2\rho, m+1+2\rho&;&\\
1, m, m+\rho, m +2\rho,m +2\rho, m +\rho +\half &;&  {|z|^2/ 4}\ea \right)  \\                          
   _7{\cal F}_8(|z|^2; m, \rho) & = &  _7F_8\left(\ba{lcr}  m+1,m+1,m+1,2m+2\rho,2m+2\rho,m+1+2\rho, m+1+2\rho&;&\\
 1, m, m,m+\rho,  m+2\rho, m+2\rho, m+2\rho, m+\rho+\half&;&  {|z|^2/ 4} \ea\right).                          
\enano}
where we adopt in the sequel the following notation 
\bea\label{pfqlong}
_pF_q\left(\ba{lcr} a_1, \ldots, a_p  &;&\\ 
                    b_1, \ldots, b_q  &;& x\ea\right).
\ena
for long hypergometric function instead of the form in  (\ref{eq59}).
The PND (\ref{PND}) reads as 
{\small
\bea\label{PND_Atype1}
{\cal P}_n^{(m)}(|z|^2;  \rho) = {\Gamma(m +2\rho)\Gamma(n+1) \Gamma(n + m + 2\rho)^2 \over \Gamma(2m + 2\rho)
\Gamma(m+1) \Gamma(n +2\rho) \Gamma(2n +2\rho)\,  _3{\cal F}_4 (|z|^2; m, \rho)}\, 
{(|z|)^{2(n-m)} \over ((n-m)!)^2}.
\ena}
\newline
\begin{figure}[htbp]
\begin{center}
\begin{minipage}{.45\textwidth}
\includegraphics[width=7cm]{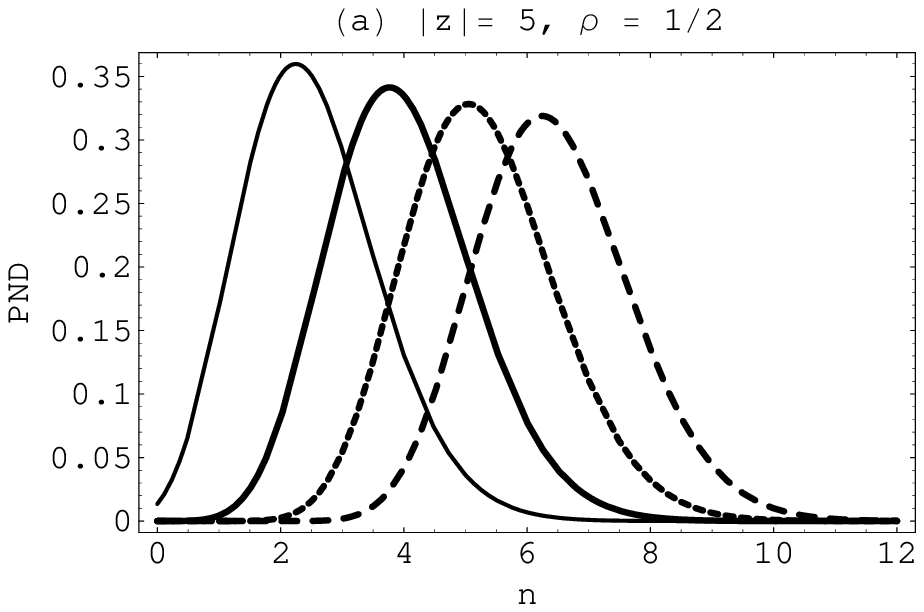}
\end{minipage} 
 \begin{minipage}{.45\textwidth}
\includegraphics[width=7cm]{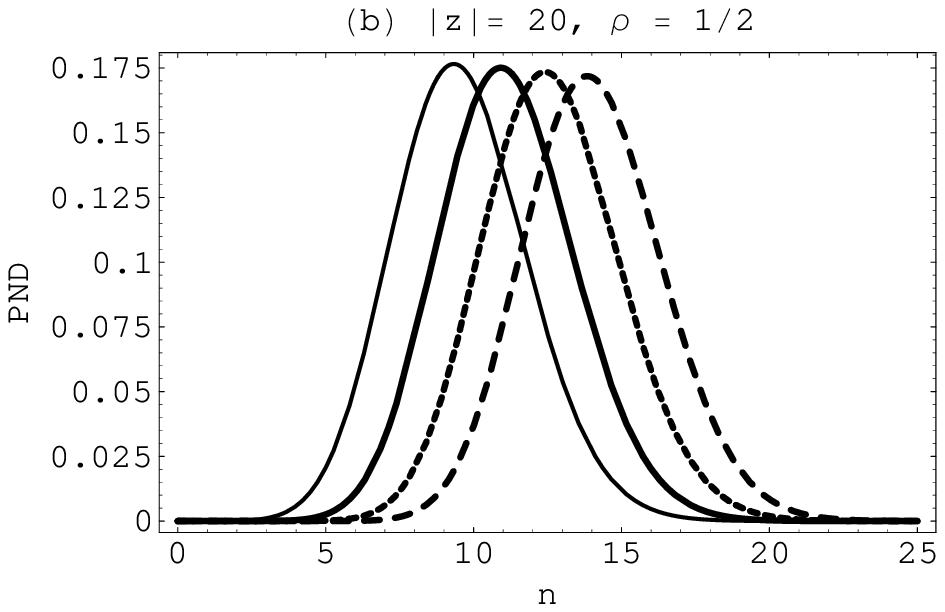}
\end{minipage}  
\end{center}
\ni \caption[]{ Plots of the  photon number dstribution  (\ref{PND_Atype1}) of the
  PA-SIPCS (\ref{eq99}) versus  the photon number $n$  for different values of the
photon-added number $m$  with $\rho = \half$ as parameter  for  $|z| = 5$(a) and  $|z| = 20$(b),   respectively, with 
$m= 0$ (thin solid  line), 
$m = 1$ (solid line), $m = 2$ (dot line) and $m = 3$ (dashed line). }
\end{figure}
\ni {
In FIG. 8,  the Mandel Q-parameter (\ref{QM_Atype1}) and the second-order correlation
function (\ref{g2_Atype1}) calculated in the   PA-SIPCS (\ref{eq99}) are plotted in terms of the amplitude $|z|$, 
in (a) and (b), respectively, for different values of the photon-added number $m$ 
with the potential parameter $\rho = \half$. 
We see that the Mandel Q-parameter increases with the amplitude $|z|$, while the second-order correlation function decreases.
In addition, for sufficiently large  values of  $|z|$, the Mandel Q-parameter  is positive while  the second-order correlation function is greater than one. 
Therefore, the   PA-SIPCS (\ref{eq99}) exhibit super-Poissonian distribution for large values of $|z|$.\\ 
In 
FIG. 9,   the PND derived in  the  PA-SIPCS (\ref{eq76})
 as a function   of $|z|$, is depicted for  $|z| = 5$ and  $|z| = 20$, in  (a) and 
 (b) with  the potential parameter $\rho = \half$, respectively, for different values of the photon-added  number $m$.  
 The  peaks position increases with
 both the photon-added number  $m$ and  the amplitude $|z|$. }
 \subsubsection{Second  choice of the functional ${\cal Z}_j$}
\ni We make  the second choice  of the functional ${\cal Z}_j$ as:
\bea \label{eq109}
{\cal Z}_j = \sqrt{g(a_1; -2\kappa, \kappa)g(a_1; -2\kappa, 2\kappa) } \, e^{- i\alpha {\cal R}}(a_1)
\ena
with $\kappa$ a real constant and where we use the auxiliary function (\ref{eq72}).
 From the potential parameter relations (\ref{eq50}) and Eq. (\ref{eq73}) we obtain:  
\bea \label{eq110}
\prod_{k = m}^{n+m-1} {\cal Z}_{j + k}  = \sqrt{\kappa^{2n}
 {\Gamma(2n + 2 m + \nu + 1) \over \Gamma ( 2 m + \nu + 1) }}e^{-i \alpha E_n}
\ena
where we assume $a_1 = -{\nu\over 2},$  the eigenenergies   being  $E_n = \kappa^2 n(n+ \nu  +1)$.  
Inserting Eqs. (\ref{eq110}), (\ref{eq94}) and (\ref{eq95}) in the expansion coefficient (\ref{eq28}), we obtain  
\bea \label{eq111}
K_n^m(a_r) = \left[ {1\over \kappa^{2m}}{ \Gamma(n + 1)^2\, \Gamma(n + m + \nu + 1)\Gamma( 2 m + \nu + 1) \over \Gamma(n + m + 1)\Gamma(n +2 m +\nu + 1)^2 
 }\right]^{\half}e^{i \alpha E_n}, 
\ena
where we assume $\beta = \kappa$ and $\rho = {\nu\over 2} +  \half$ in (\ref{eq94}) and (\ref{eq95}). 
For $m = 0$, we recover the coefficient $h_n$ in \cite{Aleixo}:
\bea \label{eq112}
K_n^0(a_r) = \left[{\Gamma(n+1) \Gamma(\nu +1) \over\Gamma(n +\nu+1)}\right]^{\half} e^{i \alpha E_n} = h_n(a_r)\pt
\ena
 {\it (i) Normalization} \\
 The normalization factor in terms of hypergeometric and Meijer's G-functions is 
{\small
\bea \label{eq113}
 {\cal N}_m(|z|^2; a_r) & = & \xi(m,\nu)\, 
 \left[{}_3F_2\left(\ba{lcr}m+1, 2m+\nu +1, 2m+\nu +1 &;&\\
 1, m+\nu +1& ;& |z|^2\ea\right)\right]^{-\half} \\ \label{eq114}
 {\cal N}_m(|z|^2; a_r)& = &
\left[{ \kappa^{2m} \over \Gamma(2m+\nu + 1)} G_{3,3}^{1,3}\left(-|z|^2\left|
\ba{ccc}-m, -2m-\nu, -2m-\nu   \quad  \\ 0 , 0, - m - \nu \ea \right.\right)\right]^{-\half}
\ena}
where $\xi(m, \nu) = \kappa^{2m} \Gamma(m+1){\Gamma(2m+ \nu+1) \over \Gamma(m + \nu +1)} $. 
For $m=0$, we recover  the normalization factor 
\bea \label{eq115}
{\cal N}_0(|z|^2; a_r) = {}_1F_0(\nu + 1;-; |z|^2)^{-\half}  = (1 - |z|^2)^{-1/2 - \nu/2 } =
 {\cal N}(|z|^2; a_r)
\ena
obtained in \cite{Aleixo}. In this case, the explicit form of the PA-SIPCS,  defined for $|z| < 1$, 
 is provided by: 
{\footnotesize
\bea \label{eq116}
 \lv z; a_r\rs_m = {\cal N}_m (|z|^2; a_r) \sum_{n = 0}^\infty 
\sqrt{ \kappa^{2m}{\Gamma(n + m + 1)\Gamma(n +2 m +\nu + 1)^2 \over 
\Gamma(n + 1)^2 \Gamma(n + m +\nu + 1)  \Gamma ( 2 m + \nu + 1) }}z^{n}e^{-i \alpha E_n} \lv n + m\rs.
\ena}
\begin{figure}[htbp]
\begin{center}
\includegraphics[width=8cm]{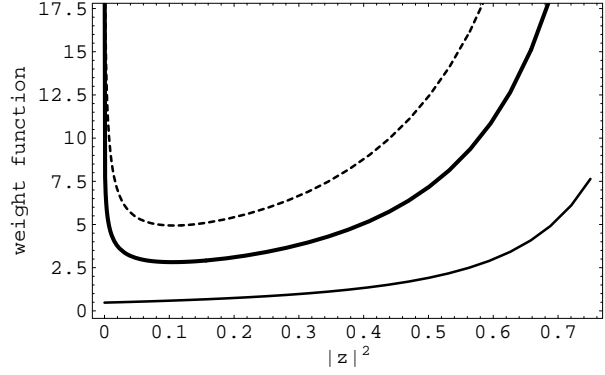}
\end{center}
\ni \caption[]{
Plots of the  measure (\ref{eq117}) of  the
 PA-SIPCS (\ref{eq116}) versus  $|z|^2$ for different  values of 
 the photon added number $m$  with 
 $m = 0$ (thin solid  line), $m = 1$ (solid line) and  $m = 2$ (dot line)}
\end{figure}
 {\it (ii) Non-orthogonality} \\
 The inner product of two different PA-SIPCS $\lv z; a_r\rs_m$ and $\lv z';a_r\rs_{m'}$
 is given by:
{\footnotesize 
\beano
 _{m'}\li z';a_r\rv \left. z; a_r\rs_m & = & \chi(z',z, m, m',\nu)\, 
_3F_2\left(\ba{lcr}m +1, m+ m' +\nu+1, 2m+\nu+1 &; &\\
m - m'+ 1, m+\nu+1 &; &{ z'}^\star z \ea\right)
\enano}
where 
{\small 
\beano
 \chi(z',z, m, m',\nu') & = &{\cal N}_{m'}(|z'|^2;a_r)\ {\cal N}_m(|z|^2;a_r) 
\ { {{ z'}^\star}^{(m-m')} \kappa^{(m+m')} \over \sqrt{\Gamma(2m +\nu+1) \Gamma(2m'+\nu +1) }}\times \\
 &  & \times 
 {\Gamma(m +1)  \Gamma(m +m'+\nu+1)  \Gamma(2m +\nu+1) \over \Gamma(m -m'+1)  \Gamma(m +\nu+1)}
\ e^ {i\alpha(E_n-E_{n+m-m'})}.
\enano }
{\it (iii) Overcompleteness}\\
\ni Following the steps and the method of the previous subsections,  we obtain the weight-function of the  PA-SIPCS  as 
{
{\footnotesize
\bea \label{eq117}
\omega_m(|z|^2; a_r) & = &{1 \over \pi} G_{3,3}^{1,3}\left(-|z|^2\left|
\ba{l}-m, -2m-\nu, -2m-\nu   \quad  \\ 0 , 0, - m - \nu \ea \right.\right)\,
G_{3,3}^{3,0}\left(|z|^2\left|
\ba{l} m, 2 m + \nu, 2 m + \nu  \\0, 0, m+ \nu   \ea \right.\right) 
\ena}
 The measure (\ref{eq117}) as  represented in FIG. 10, for $\nu = 1.5
$ and $m = 0, 1, 2$ is  positive for $\nu > 0 $.} 
We recover, for $m = 0$, the result:
\bea \label{eq118}
\omega_0(|z|^2; a_r) & = & {\Gamma(\nu + 1) \over \pi}\, 
_1F_0(\nu + 1;-; |z|^2)\, 
G_{1,1}^{1,0}\left(|z|^2\left|
\ba{ccc} \quad & ; & \nu  \\0  & ; &   \ea \right.\right) = {\nu \over \pi} (1 - |z|^2)^{-2}
\ena
obtained in \cite{Aleixo} for the corresponding ordinary SIPCS. 
\\
 {\it (iv) Statistical properties} \\
\begin{figure}[htbp]
\begin{center}
\begin{minipage}{.45\textwidth}
\includegraphics[width=8cm]{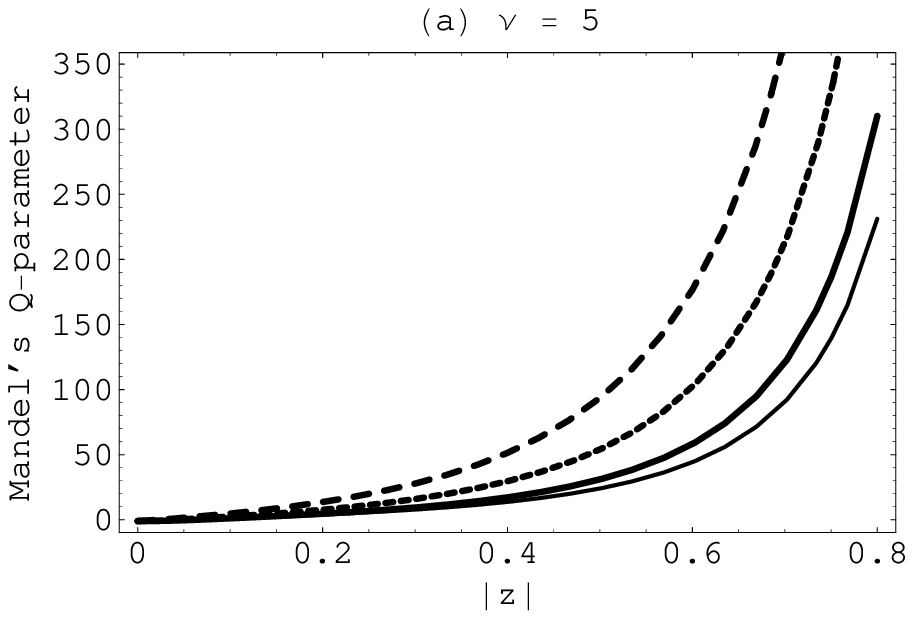}
\end{minipage} 
 \begin{minipage}{.45\textwidth}
\includegraphics[width=8cm]{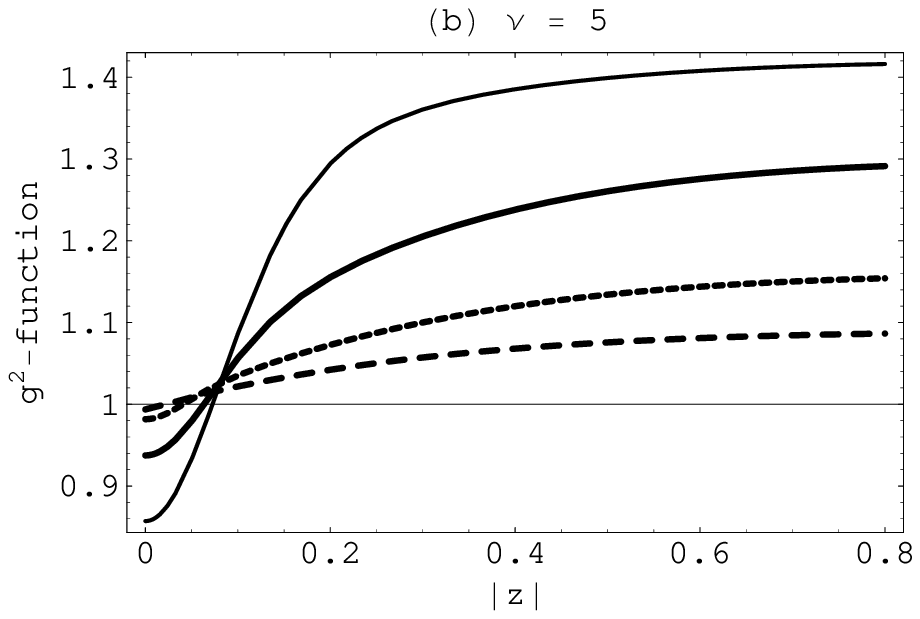}
\end{minipage}  
\end{center} 
\ni \caption[]{ Plots of the  Mandel Q-parameter  (\ref{QM_Atype2})(a)   and the second-order correlation function (\ref{g2_Atype2})(b)  of the
 PA-SIPCS (\ref{eq116}) versus   $|z|$  with the parameter $ \nu = 5$   for different values of 
 the photon added number $m$ with 
  $m = 1$ (thin solid  line), $m = 2$ (solid line), $m = 5$ (dot line), 
and $m = 10$ (dashed line).}
\end{figure}
The expectation values $\li N\rs$ and $\li N^2\rs$  in the state (\ref{eq116})
are provided as follows:
{\footnotesize
\bea \label{eq119} 
\li N\rs  & = &  \kappa^2 m(m+ \nu + 1) {_5{\cal F}_4 (|z|^2; m, \nu) \over _3{\cal F}_2(|z|^2; m, \nu)}\ , \ 
\li N^2\rs =   \kappa^4 m^2(m+ \nu + 1)^2 {_7{\cal F}_6(|z|^2; m, \nu) \over _3{\cal F}_2(|z|^2; m, \nu)}\pt
\ena}
Then, the Mandel Q-parameter and the second-order correlation function are  given by:
\bea \label{QM_Atype2}
 Q = \kappa^2 m (m + \nu + 1) \left( {_7{\cal F}_6(|z|^2;  m, \nu)\over _5{\cal F}_4 (|z|^2;  m, \nu)} -
     { _5{\cal F}_4(|z|^2;  m, \nu)\over _3{\cal F}_2 (|z|^2;  m, \nu)} \right) - 1
\ena
\bea  \label{g2_Atype2}
 g^2 = {m \kappa^2 (m + \nu +1)\ _7{\cal F}_6(|z|^2; m, \nu) -  _5{\cal F}_4 (|z|^2; m, \nu)
 \over m \kappa^2 (m + \nu +1)\  _5{\cal F}_4 (|z|^2; m, \nu)}\, 
{ _3{\cal F}_2 (|z|^2; m, \nu) \over _5{\cal F}_4 (|z|^2;  m, \nu)}
\ena
where $_3{\cal F}_2 $, $_5{\cal F}_4$ and  $_7{\cal F}_6$ are the generalized hypergeometric functions:
{\footnotesize 
\beano
  _3{\cal F}_2 (|z|^2; m, \nu) & = & _3F_2\left(\ba{lcr}m+1,2m +\nu + 1, 2m +\nu + 1 &;&\\
  1 , m+\nu +1& ;& |z|^2\ea\right) \\
  _5{\cal F}_4 (|z|^2; m, \nu) & = &  _5F_4\left(\ba{lcr} m+1,m+1, 2m +\nu + 1, 2m +\nu + 1, m+ \nu + 2&;&\\ 
1, m, m+\nu + 1,  m+\nu + 1 &; &|z|^2\ea\right) \\
 _7{\cal F}_6(|z|^2; m, \nu) & = & _7F_6\left(\ba{lcr}m+1, m+1, m+1, 2m +\nu + 1,2m +\nu + 1, m+ \nu + 2, m+ \nu + 2 &
; & \\
 1, m, m, m+\nu + 1,m+\nu + 1,m+\nu + 1 &;&  |z|^2\ea\right).
\enano}
The PND (\ref{PND}) reads as 
{\footnotesize
\bea\label{PND_Atype2}
 {\cal P}_n^{(m)}(|z|^2;  \nu) = {\Gamma(m + \nu + 1)\Gamma(n+1) \Gamma(n + m + \nu + 1)^2 \over 
\Gamma(2m + \nu + 1)^2 \Gamma(m+1) \Gamma(n  + \nu + 1) \,  _3{\cal F}_2 (|z|^2; m, \nu)}\, 
{(|z|)^{2(n-m)} \over ((n-m)!)^2}
\ena}
\begin{figure}[htbp]
\begin{center}
\begin{minipage}{.45\textwidth}
\includegraphics[width=8.0cm]{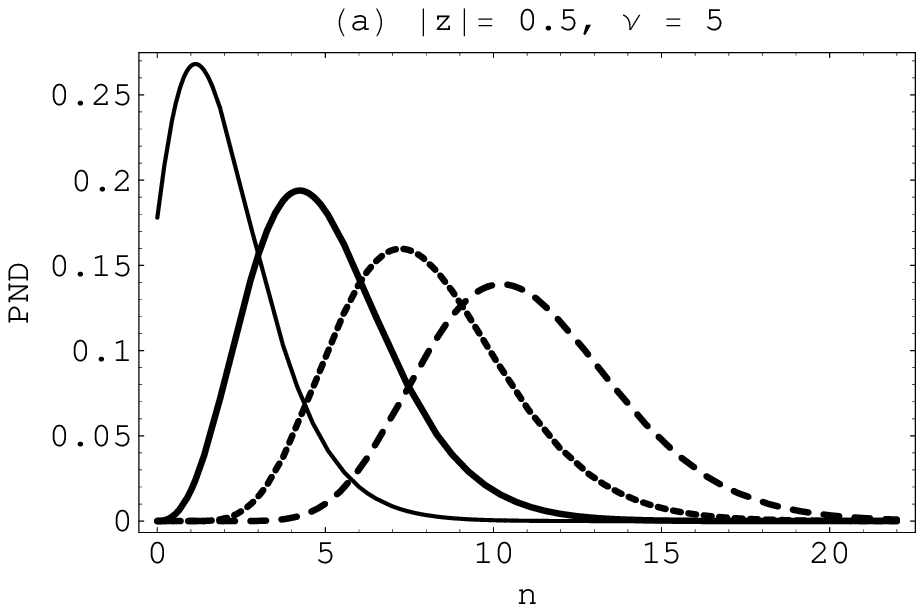}
\end{minipage} 
 \begin{minipage}{.45\textwidth}
\includegraphics[width=8.0cm]{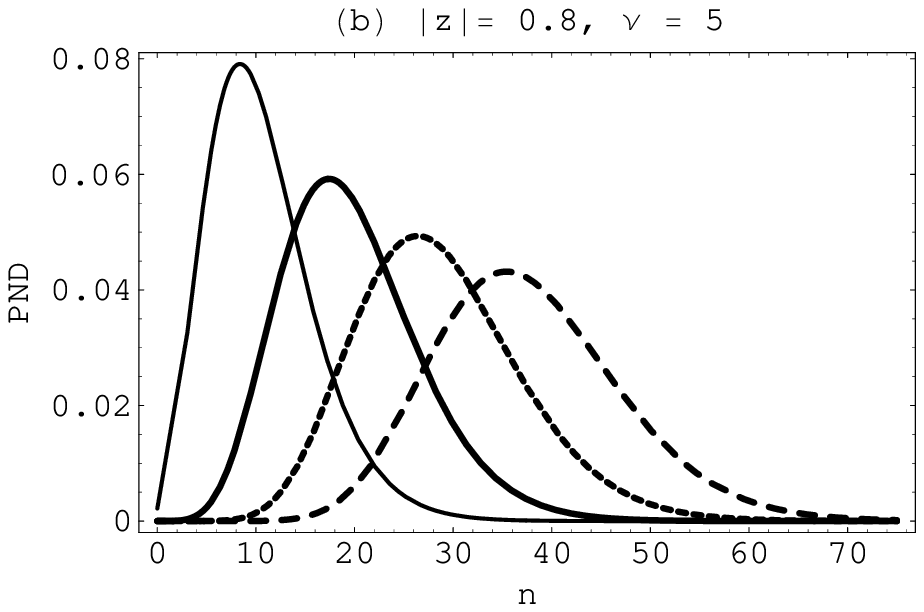}
\end{minipage}  
\end{center} 
\ni \caption[]{ Plots of the  photon number dstribution  (\ref{PND_Atype2}) of the
  PA-SIPCS (\ref{eq116}) versus the photon number $n$  for different values of the
photon-added number $m$ with the  parameter  $\nu = 5$   for  $|z| = 0.5$(a)  and  $|z| = 0.8$ (b),   respectively, with 
$m= 0$ (thin solid  line), $m = 1$ (solid line), $m = 2$ (dot line) and $m = 3$ (dashed line). }
\end{figure}
\ni{In FIG. 11,    the Mandel Q-parameter (\ref{QM_Atype2}) and the second-order correlation
function (\ref{g2_Atype2}) calculated in the   PA-SIPCS (\ref{eq116}) are plotted in terms of the amplitude $|z|$, 
in (a) and (b), respectively, for different values of the photon-added number $m$ 
with the potential parameter $\nu = 5$.   
 We observe that the Mandel Q-parameter increases with the amplitude $|z|$, while the second-order correlation function decreases. Furthermore, for sufficiently large values of $|z|$, the 
 Mandel Q-parameter  is  positive while the second-order correlation function is greater than one. 
Therefore, the   PA-SIPCS (\ref{eq116}) exhibit super-Poissonian distribution for large values of $|z|$.
 
In FIG. 12,   the PND derived in  the  PA-SIPCS  (\ref{eq116})   as a function   of $|z|$, is depicted for  $|z| = 0.5$ and  $|z| = 0.8$, in  (a) and 
 (b) with  the potential parameter $\nu = 5$, respectively, for different values of the photon-added  number $m$.   Increasing  the photon-added number  $m$ engenders a clear shift of  the PND.  
 Comparing Figures (a) and (b), it comes that the peaks position increases with  the amplitude $|z|$.}


\section{Conclusion}
\ni In this paper,  a set  of non-classical states, i.e, the photon-added coherent states   
associated with the shape-invariant systems, denoted PA-SIPCS,   have been constructed and fully characterized from mathematics  
and physics points of view.
The formalism has been illustrated  on some relevant examples withdrawn from Infeld and Hull \cite{Infeld} classes.
The generalized states obtained here encompass the previous known results in the literature as particular cases.
Relevant physical quantities have been expressed in terms of generalized hypergeometric functions $_pF_q$ and the Meijer's  
G-functions. The moment problem has  been explicitly   solved by using the Mellin inversion theorem and the Meijer's  G-function.\\
The statistical properties involving the PND, Q-Mandel parameter  and the second-order correlation function  of the investigated physical systems have been described and thoroughly  
discussed. The Poissonian, sub-Poissonian and super-Poissonian distribution behaviors of the PA-SIPCS in the studied  main physical systems 
have been highlighted. 


\begin{thebibliography}{32}
\bibitem{Schrodinger}{E. Schr\"odinger,   {Naturwiss}\, {\bf 14}, 664 (1926)}
\bibitem{Klauder_App}{ J. R. Klauder, B.-S. Skagerdtam,   {\it Coherent states: applications in physics and mathematical physics}, (World Sientific, Singapore, 1985); {J.  R. Klauder  \, 
  { Phys. Rev. D}  {\bf 19},   2349  (1979)}}
 \bibitem{Pere}{A. M. Perelomov,   {\it Generalized coherent states and their applications}  (Springer, Berlin, 1986)}
\bibitem{Ali95}{S. T. Ali, J. P. Antoine, J.-P. Gazeau  and U. A. Mueller,  {Rev. Math. Phys.}\, 
 {\bf 7},  1013 (1995)}
\bibitem{Gazeau}{J.-P.Gazeau, {\it Coherent states in quantum mechanics}  (Wiley-VCH,Weinheim, 2009)}
\bibitem{Berezin}{F. A. Berezin, {\it The method of Second Quantization} (Nauka, Moscow, 1986)  (1986)}
\bibitem{Aremua1}{I. Aremua, J.-P. Gazeau and M. N. Hounkonnou,  {J. Phys. A: Math. Gen}  {\bf 45},  335302 (2012) }
 \bibitem{Barut}{A. O. Barut and L. Girardello,  {Commun. Math. Phys.}  {\bf 21},  41  (1971)}
\bibitem{Perelomov}{A. M. Perelomov,   {Commun. Math. Phys.} {\bf 26}, 222 (1972)}
 \bibitem{Aragone}{C. Aragone, G. Guerri, S. Salam\'o  and  J. L. Tanin  
  {J. Phys. A: Math. Gen.} {\bf 7}, L149 (1974)}
  \bibitem{Nieto}{M. M. Nieto and L. M. Jr. Simmons,    {Phys. Rev. Lett.}  {\bf 41},  207 (1978)}
 \bibitem{Fukui}{T. Fukui  and N. Aizawa,    {Phys. Lett. A}  {\bf 189},  7 (1994)  }
 \bibitem{Aleixo}{A. N. F. Aleixo, A. B. Balantekin and  M. A. C\^andido Ribeiro,
   {J. Phys. A: Math. Gen.}  {\bf 35},  9063 (2002)}
\bibitem{Witten}{E.  Witten,  {Nucl. Phys. B} {\bf 185},   5123 (1981)}
\bibitem{Cooper}{F. Cooper, A. Khare and U.  Sukhatme,   
 {Phys. Rep.} {\bf 251},   267  (1995)}
\bibitem{Gend}{L. Gendenshtein,  {JETP Lett.} {\bf 38}, 356 (1983)  }
 \bibitem{Dutt4}{R. Dutt, A. Khare and U. Sukhatme,
   {Ame. J. Phys.}  {\bf 56}(2),  163 (1988)}
\bibitem{Dab}{J. Dabrowska, A. Khare and U. Sukhatme, {J.
Phys. A: Math. Gen.} {\bf 21}, L195 (1988)}
\bibitem{Dodonov}{V. V. Dodonov, M. A. Marchiolli, Ya. A. Korenmoy, V. I. Man'ko and Y. A. Mouchin,  \,  
{Phys. Rev. A} {\bf 58},  4087 (1998)}
\bibitem{Popov}{D. Popov,    {J. Phys A: Math. Gen}  {\bf 35},  7205 (2002)}
\bibitem{Berrada}{K. Berrada,  {J. Math. Phys} {\bf 56}, 072104 (2015)}
\bibitem{Daoud}{Daoud M,  {Phys. Lett. A.} {\bf 305}, 135 (2002)}
\bibitem{Aga}{G. S. Agarwal and  K. Tara,  {Phys. Rev A.}  {\bf 43}, 492 (1991); 
G. S. Agarwal  and K. Tara,  {phys. Rev A.} {\bf 46}, 485 (1992)}
\bibitem{Lietal}{H. M. Li, H. C. Yuan and H. Y. Fan,  {Int. J. Theor. Phys.} {\bf 48}, 
2849 (2009)}
\bibitem{Zhang}{J. S. Zhang  and J. B. Xu,  {Phys. Scr.} {\bf 79}, 025008 (2009)}
\bibitem{hounk-ngompe}{M. N. Hounkonnou and  E. B.  Ngompe Nkouankam,   {J. Phys A: Math. Theor.}
 {\bf 42}(2), (2008)}
\bibitem{mojaverietal1}{B.  Mojaveri, A. Dehghani   and S. Mahmoodi,      {Phys. Scr.}  {\bf 89},  085202 (2014)}
\bibitem{mojaverietal2}{B. Mojaveri and A. Dehghani, {Eur. Phys. J. D } {\bf  68}, 315 (2014)}
\bibitem{Dakna}{M. Dakna, T. Anhut, T. Opatrny, L. Kn\"oll and D-G. Welsch, {Phys. Rev.A.}
 {\bf 55}, 3184 (1997); 
{M. Dakna, L. Kn\"oll,   and D-G Welsch,  {Opt. Commun.} {\bf 145}, 309 (1998)}}
\bibitem{Ban}{M. Ban,  {J. Mod. Opt.} {\bf 43}, 1281 (1996)}
\bibitem{Mandel95}{L. Mandel  and E. Wolf,  {Optical coherence and quantum optics}  (Cambridge University Press, Cambridge, 1995) (1995)}
\bibitem{Penson99}{K. A. Penson and A. I. Solomon,   {J. Math. Phys.} {\bf 40}, 2354 (1999)}
\bibitem{Infeld}{L. Infeld  and  T. E. Hull,  {Rev. Mod. Phys.} {\bf 23},  28 (1951)}
\bibitem{Marichev}{O. I. Marichev,   {Handbook of integral transforms
of higher transcendental
 functions: theory and algorithmic tables}\,  (Ellis Harwood, Chichester, UK, 1983)}
 \bibitem{Ober}{F. Oberhettinger,  {\it Tables of Mellin transforms}
  (Springer, Berlin, 1974)}
\bibitem{Mathai}{A. M. Mathai and  R. K. Saxena, 
 {\it Generalized Hypergeometric Functions with Applications in Statistics and
Physical Sciences} (Lecture Notes in Mathematics vol 348) (Springer, Berlin, 1973)} 
\end{thebibliography}
\end{document}